\documentclass[twocolumn]{aastex7}
\usepackage[T1]{fontenc}
\usepackage{apjfonts} 
\usepackage[figure,figure*]{hypcap}
\usepackage{comment}
\usepackage{amssymb}
\usepackage{amsmath}
\usepackage{longtable}
\usepackage[T1]{tipa}
\usepackage{CJKutf8}
\usepackage{gensymb}

\newcommand{\cat}{Catalog 2}
\newcommand{\frb}{FRB 20210705}
\newcommand{\halpha}{H$\alpha$}
\newcommand{\fitburst}[0]{{\tt fitburst}}

\newcommand{\dmu}{$\text{ pc cm}^{-3}$}
\newcommand{\emu}{$\text{ pc cm}^{-6}$}
\newcommand{\cmc}{$\text{ cm}^{-3}$}
\newcommand{\numfrbs}{3542}
\newcommand{\numnonrepfrbs}{3459}
\newcommand{\numrepfrbs}{83}

\defcitealias{chimefrbcatalog1}{CHIME/FRB Collaboration et~al.~2021}
\defcitealias{abb+18}{CHIME/FRB Collaboration et~al.~2018}
\defcitealias{chimefrbcatalog2}{CHIME/FRB Collaboration et~al.~2025}
\defcitealias{abb+21}{CHIME/Pulsar Collaboration et~al.~2021}
\defcitealias{abb+22}{CHIME Collaboration et~al.~2022}
\defcitealias{aaa+25_published}{CHIME Collaboration et~al.~2025}
\defcitealias{abb+19c}{CHIME/FRB Collaboration et~al.~2019}
\defcitealias{planckforegrounds16}{Planck Collaboration et~al.~2016}
\defcitealias{CHAMPSS+25}{The CHAMPSS Collaboration et~al.~2025}

\begin{document}

\shortauthors{Patil et al.}

\shorttitle{A Spatial Gap in the FRB Sky Distribution}

\title{A Spatial Gap in the Sky Distribution of Fast Radio Burst Detections Coinciding with Galactic Plasma Overdensities}

\newcommand{\ICRAR}{\affiliation{International Centre for Radio Astronomy Research (ICRAR), Curtin University, Bentley WA 6102 Australia}}

\newcommand{\DRAO}{\affiliation{Dominion Radio Astrophysical Observatory, Herzberg Research Centre for Astronomy and Astrophysics, National Research Council Canada, PO Box 248, Penticton, BC V2A 6J9, Canada}}

\newcommand{\CSIRO}{\affiliation{CSIRO Space \& Astronomy, Parkes Observatory, P.O. Box 276, Parkes NSW 2870, Australia}}

\newcommand{\NRC}{\affiliation{NRC Herzberg Astronomy and Astrophysics, 5071 West Saanich Road, Victoria, BC V9E2E7, Canada}}

\newcommand{\CIERA}{\affiliation{Center for Interdisciplinary Exploration and Research in Astronomy (CIERA), Northwestern University, 1800 Sherman Avenue, Evanston, IL 60201, USA }}

\newcommand{\NU}{\affiliation{Department of Physics and Astronomy, Northwestern University, Evanston, IL 60208, USA}}

\newcommand{\Uch}
{\affiliation{Department of Astronomy and Astrophysics, University of Chicago, William Eckhart Research Center, 5640 South Ellis Avenue, Chicago, IL 60637, USA}}

\newcommand{\UCSC}{\affiliation{Department of Astronomy and Astrophysics, University of California, Santa Cruz, 1156 High Street, Santa Cruz, CA 95064, USA}}

\newcommand{\IPMU}{\affiliation{Kavli Institute for the Physics and Mathematics of the Universe (Kavli IPMU), 5-1-5 Kashiwanoha, Kashiwa, 277-8583, Japan}}

\newcommand{\NAOJ}{\affiliation{Division of Science, National Astronomical Observatory of Japan, 2-21-1 Osawa, Mitaka, Tokyo 181-8588, Japan}}

\newcommand{\MU}{\affiliation{Department of Physics, McGill University, 3600 rue University, Montr\'eal, QC H3A 2T8, Canada}}

\newcommand{\TSI}{\affiliation{Trottier Space Institute, McGill University, 3550 rue University, Montr\'eal, QC H3A 2A7, Canada}}

\newcommand{\CMU}{\affiliation{McWilliams Center for Cosmology \& Astrophysics, Department of Physics, Carnegie Mellon University, Pittsburgh, PA 15213, USA}}

\newcommand{\UVA}
{\affiliation{Anton Pannekoek Institute for Astronomy, University of Amsterdam, Science Park 904, 1098 XH Amsterdam, The Netherlands}}

\newcommand{\ASTRON}
{\affiliation{ASTRON, Netherlands Institute for Radio Astronomy, Oude Hoogeveensedijk 4, 7991 PD Dwingeloo, The Netherlands
}}

\newcommand{\MITK}
{\affiliation{MIT Kavli Institute for Astrophysics and Space Research, Massachusetts Institute of Technology, 77 Massachusetts Ave, Cambridge, MA 02139, USA}}

\newcommand{\MITP}
{\affiliation{Department of Physics, Massachusetts Institute of Technology, 77 Massachusetts Ave, Cambridge, MA 02139, USA}}

\newcommand{\CCAPS}{\affiliation{Cornell Center for Astrophysics and Planetary Science, Cornell University, Ithaca, NY 14853, USA}}

\newcommand{\DI}
{\affiliation{Dunlap Institute for Astronomy and Astrophysics, 50 St. George Street, University of Toronto, ON M5S 3H4, Canada}}

\newcommand{\DAA}
{\affiliation{David A. Dunlap Department of Astronomy and Astrophysics, 50 St. George Street, University of Toronto, ON M5S 3H4, Canada}}

\newcommand{\STSCI}
{\affiliation{Space Telescope Science Institute, 3700 San Martin Drive, Baltimore, MD 21218, USA}}

\newcommand{\WVUPA}
{\affiliation{Department of Physics and Astronomy, West Virginia University, PO Box 6315, Morgantown, WV 26506, USA }}

\newcommand{\WVUGWAC}
{\affiliation{Center for Gravitational Waves and Cosmology, West Virginia University, Chestnut Ridge Research Building, Morgantown, WV 26505, USA}}

\newcommand{\UCBASTRO}
{\affiliation{Department of Astronomy, University of California, Berkeley, CA 94720, United States}}

\newcommand{\YORK}
{\affiliation{Department of Physics and Astronomy, York University, 4700 Keele Street, Toronto, ON MJ3 1P3, Canada}}

\newcommand{\PI}
{\affiliation{Perimeter Institute of Theoretical Physics, 31 Caroline Street North, Waterloo, ON N2L 2Y5, Canada}}

\newcommand{\UBC}
{\affiliation{Department of Physics and Astronomy, University of British Columbia, 6224 Agricultural Road, Vancouver, BC V6T 1Z1 Canada}}

\newcommand{\UCHILE}
{\affiliation{Department of Electrical Engineering, Universidad de Chile, Av. Tupper 2007, Santiago 8370451, Chile}}

\newcommand{\MIBR}
{\affiliation{Miller Institute for Basic Research, University of California, Berkeley, CA 94720, United States}}

\newcommand{\UDOM}
{\affiliation{Department of Physics, College of Natural and Mathematical Sciences, University of Dodoma, 1 Benjamin Mkapa Road, 41218 Iyumbu, Dodoma 259, Tanzania}}

\newcommand{\DoSS}
{\affiliation{Department of Statistical Science, University of Toronto, Ontario Power Building, 700 University Avenue, 9th Floor, Toronto, ON M5G 1Z5, Toronto, Ontario, Canada}}

\newcommand{\IITK}
{\affiliation{Indian Institute Of Technology Kanpur, Kanpur, Uttar Pradesh 208016, India}}
\author[0009-0008-7264-1778]{Swarali Shivraj Patil}
  \WVUPA
  \WVUGWAC
  \email{sp00049@mix.wvu.edu}
\author[0000-0002-7164-9507]{Robert A.~Main}
  \MU
  \TSI
  \email{robert.main@mcgill.ca}
\author[0000-0001-8384-5049]{Emmanuel Fonseca}
  \WVUPA
  \WVUGWAC
  \email{emmanuel.fonseca@mail.wvu.edu}
\author[0000-0003-2111-3437]{Kyle McGregor}
  \MU
  \TSI
  \email{kyle.mcgregor@mail.mcgill.ca}
\author[0000-0002-3382-9558]{B.~M.~Gaensler}
  \UCSC
  \DI
  \DAA
  \email{gaensler@ucsc.edu}
\author[0000-0002-3615-3514]{Mohit Bhardwaj}
  \IITK
  \WVUPA
  \email{mohitb@iitk.ac.in}
\author[0000-0002-1800-8233]{Charanjot Brar}
  \NRC
  \email{charanjot.brar@nrc-cnrc.gc.ca}
\author[0000-0001-6422-8125]{Amanda M. Cook}
  \MU
  \TSI
  \UVA
  \email{amanda.cook@mail.mcgill.ca}
\author[0000-0002-8376-1563]{Alice P.~Curtin}
  \MU
  \TSI
  \email{alice.curtin@mail.mcgill.ca}
\author[0000-0003-3734-8177]{Gwendolyn Eadie}
  \DAA
  \DoSS
  \email{gwen.eadie@utoronto.ca}
\author[0000-0003-3457-4670]{Ronniy Joseph}
  \MU
  \TSI
  \email{ronniy.joseph@mcgill.ca}
\author[0009-0007-5296-4046]{Lordrick Kahinga}
  \UCSC
  \UDOM
  \email{lkahinga@ucsc.edu}
\author[0000-0001-9345-0307]{Victoria Kaspi}
  \MU
  \TSI
  \email{victoria.kaspi@mcgill.ca}
\author[0009-0004-4176-0062]{Afrokk Khan}
  \MU
  \TSI
  \email{afrasiyab.khan@mcgill.ca}
\author[0009-0008-6166-1095]{Bikash Kharel}
  \WVUPA
  \WVUGWAC
  \email{bk0055@mix.wvu.edu}
\author[0000-0003-2116-3573]{Adam E.~Lanman}
  \MITK
  \MITP
  \email{alanman@mit.edu}
\author[0000-0002-4209-7408]{Calvin Leung}
  \MIBR
  \UCBASTRO
  \email{calvin_leung@berkeley.edu}
\author[0000-0002-4279-6946]{Kiyoshi W.~Masui}
  \MITK
  \MITP
  \email{kmasui@mit.edu}
\author[0000-0002-0940-6563]{Mason Ng}
  \MU
  \TSI
  \email{mason.mg@mcgill.ca}
\author[0000-0003-0510-0740]{Kenzie Nimmo}
  \MITK
  \email{knimmo@mit.edu}
\author[0000-0002-8897-1973]{Ayush Pandhi}
  \DAA
  \DI
  \email{ayush.pandhi@mail.utoronto.ca}
\author[0000-0002-8912-0732]{Aaron B.~Pearlman}
  \MU
  \TSI
  \email{aaron.b.pearlman@physics.mcgill.ca}
\author[0000-0002-4795-697X]{Ziggy Pleunis}
  \UVA
  \ASTRON
  \email{z.pleunis@uva.nl}
\author[0000-0002-4623-5329]{Mawson W.~Sammons}
  \MU
  \TSI
  \email{mawson.sammons@mcgill.ca}
\author[0000-0003-3154-3676]{Ketan R.~Sand}
  \MU
  \TSI
  \email{ketan.sand@mail.mcgill.ca}
\author[0000-0002-7374-7119]{Paul Scholz}
  \YORK
  \DI
  \email{pscholz@yorku.ca}
\author[0000-0002-6823-2073]{Kaitlyn Shin}
  \MITK
  \MITP
  \email{kshin@mit.edu}
\author[0000-0003-2631-6217]{Seth R.~Siegel}
  \PI
  \MU
  \TSI
  \email{ssiegel@perimeterinstitute.ca}
\author[0000-0002-2088-3125]{Kendrick Smith}
  \PI
  \email{kmsmith@perimeterinstitute.ca}

\newcommand{\allacks}{
  We are grateful to the anonymous referee for their constructive comments that have improved our work. We thank Loren Anderson, Liam Connor, Duncan Lorimer, Namir Kassim, and Jackson Taylor for thoughtful questions and discussions that informed the analyses presented in this paper.

  We acknowledge that CHIME is located on the traditional, ancestral, and unceded territory of the Syilx/Okanagan people. We are grateful to the staff of the Dominion Radio Astrophysical Observatory, which is operated by the National Research Council of Canada. CHIME operations are funded by a grant from the NSERC Alliance Program and by support from McGill University, University of British Columbia, and University of Toronto. CHIME was funded by a grant from the Canada Foundation for Innovation (CFI) 2012 Leading Edge Fund (Project 31170) and by contributions from the provinces of British Columbia, Québec and Ontario. The CHIME/FRB Project was funded by a grant from the CFI 2015 Innovation Fund (Project 33213) and by contributions from the provinces of British Columbia and Québec, and by the Dunlap Institute for Astronomy and Astrophysics at the University of Toronto. Additional support was provided by the Canadian Institute for Advanced Research (CIFAR), the Trottier Space Institute at McGill University, and the University of British Columbia (UBC). This research has made use of the VizieR catalogue access tool, CDS, Strasbourg, France (DOI : 10.26093/cds/vizier). The original description of the VizieR service was published in 2000, A\&AS 143, 23.  

E.F. and S.S.P. are supported by the National Science Foundation under grant AST-2407399.
K.T.M is supported by a FRQNT Master's Research Scholarship.
A.M.C. acknowledges funding from NSERC as a Banting Postdoctoral Fellow. 
A.P.C. is a Vanier Canada Graduate Scholar.
G.M.E. acknowledges support from NSERC Discovery Grant RGPIN-2020-04554.
V.M.K. holds the Lorne Trottier Chair in Astrophysics \& Cosmology, a Distinguished James McGill Professorship, and receives support from an NSERC Discovery grant (RGPIN 228738-13).
C. L. acknowledges support from the Miller Institute for Basic Research at UC Berkeley.
K.W.M. holds the Adam J. Burgasser Chair in Astrophysics and is supported by NSF grant 2018490.
M.N. is a Fonds de Recherche du Quebec -- Nature et Technologies~(FRQNT) postdoctoral fellow.
K.N. is an MIT Kavli Fellow.
A.P. is funded by the NSERC Canada Graduate Scholarships -- Doctoral program.
A.B.P. is a Banting Fellow, a McGill Space Institute~(MSI) Fellow, and a FRQNT postdoctoral fellow.
Z.P. is supported by an NWO Veni fellowship (VI.Veni.222.295).
M.W.S. acknowledges support from the Trottier Space Institute Fellowship program.
K.R.S acknowledges support from FRQNT doctoral research award.
P.S. acknowledges the support of an NSERC Discovery Grant (RGPIN-2024-06266).
K.S. is supported by the NSF Graduate Research Fellowship Program.
}

\correspondingauthor{Swarali Shivraj Patil}
\email{sp00049@mix.wvu.edu}

\begin{abstract}
    We analyze the positional and morphological properties of about 3600 unique fast radio burst (FRB) sources reported in the second FRB catalog generated by the Canadian Hydrogen Intensity Mapping Experiment (CHIME) telescope. We find a two-dimensional dependence of FRB detections on sky position, and identify a significant absence of detections in a roughly circular region centered at Galactic coordinates (77.7$\degree$, 0.9$\degree$), spanning an area of 213.6 deg$^2$. This detection gap spatially coincides with the Cygnus X region -- a plasma-rich star-forming region in the Milky Way. This feature is most likely the result of increased sky temperature and strong multi-path scattering by turbulent ionized plasma, which broadens the FRB signals beyond detectability in the CHIME band. Our simulations yield a mean of 6 expected FRB detections within the gap when accounting for the elevated sky temperature in the direction of the detection gap. We infer that a lower limit of the maximum scattering timescale $\tau_{\rm sc,\, 1\,GHz} \geq 5.59$ ms, obtained without assuming a model of the Galactic electron distribution, is sufficient to suppress the brightness of all coincident FRBs.  A similar suppression is seen in \cat{} along other high-emission measure (EM) sightlines ( i.e., EM$\geq$2900\emu), further supporting a broader trend of suppression due to Galactic scattering. Future very long baseline interferometry (VLBI) measurements of scattering disks with CHIME Outriggers can help confirm our interpretation. Our work highlights the notion that FRBs can serve as new, model-independent tracers of the warm ionized medium within our Milky Way Galaxy. 
\end{abstract}

\keywords{Radio bursts (1339) --- Radio transient sources (2008) --- Interstellar scattering (854) --- Interstellar plasma (851) --- High energy astrophysics (739)}

\section{Introduction} \label{sec:intro}

Fast radio bursts (FRBs) are bright and brief extragalactic bursts of energy emitted in the radio spectrum (\citealp{lbm+07}, also see \citealp{petroff_frb_overview} for a recent overview). While their origins remain unclear, there are a growing number of observations that support models based on compact objects, such as highly magnetized neutron stars (e.g.{\citealp{chime_frb_galactic_magnetar, brb+20}}, see \citealp{pww+19} for a collection of proposed models). Regardless of their origins, FRBs undergo the same dispersive and scatter-broadening effects regularly observed in Galactic radio pulsars \citep{occ_2022}, with scattering arising from dense ionized regions in  near-source environments within their host galaxies and/or and the Milky Way. Scattering often needs to be modeled with one or more distinct scattering screens -- a framework developed for pulsars \citep{Williamson_mnras} and subsequently applied to FRBs. In this sense, FRBs can be used as powerful probes of the electron density fluctuations in the interstellar medium (ISM) along their lines of sight.

Along a given line of sight (LOS), the scattering timescale ($\tau_{\rm sc}$) characterizes the time delay caused by multi-path propagation through the  warm ionized medium (WIM) of the Milky Way Galaxy. Propagation through ionized plasma also results in a frequency-dependent dispersive delay that is proportional to the integrated column density of free electrons along the LOS. This integrated quantity is known as the dispersion measure (DM). Based on the DM and $\tau_{\rm sc}$ predictions from the NE2001 Galactic electron density model \citep{ne2001}, \cite{occ_2022} postulated  a ``zone of avoidance" to span over a range of Galactic longitudes $|l| \leq 50\degree$ and latitudes $|b| \leq 4.1\degree$ at 0.4 GHz for FRBs. \cite{mj15} proposed that diffractive scintillation enhances FRB detectability at higher Galactic latitudes by boosting some bursts above the detection threshold, particularly in the presence of a steep luminosity distribution. Strong scattering in the Galactic plane can suppress scintillation and broaden the pulse, reducing the likelihood of such boosts. 

FRBs are therefore understood to encode information about the WIM through measurements of $\tau_{\rm sc}$, serving as ``backlights" to the foreground that is the Milky Way Galaxy. A large sample of $\tau_{\rm sc}$ measured from FRBs observed across a large fraction of the sky are expected to provide the means for constraining structural information of the WIM \citep[e.g.,][]{mj15}. However, past studies of the FRB sky distribution have been limited by the small number of detections and the challenges associated with combining data from different instruments, each with differing sensitivities, instrumental uncertainties, and selection functions. The FRB sky distribution can greatly vary based on the observing frequency. As a result, the scope of such studies so far has been limited to probing only the Galactic latitude dependence \citep{phl19}. Using data acquired with the Canadian Hydrogen Intensity Mapping Experiment (CHIME) Fast Radio Burst project (hereafter, CHIME/FRB), \cite{jcc+21} reported no significant dependence of FRB detections on Galactic latitude in CHIME/FRB Catalog 1 (\citetalias{chimefrbcatalog1}) containing 536 FRBs. 

The forthcoming second catalog of CHIME/FRB (hereafter \cat{}) represents a significant advancement, reporting the largest number of unique FRBs detected by a single instrument (\citetalias{chimefrbcatalog2}). An updated map of the FRB sky positions in \cat{} reveals a striking and visually discernible gap in FRB detections. For the first time, we clearly identify a Galactic scattering zone of avoidance (hereafter referred to as the ``detection gap'') in FRB detections, which spans Galactic longitudes $ 70\degree < l < 90\degree$ and latitudes $ -7\degree < b < 11\degree$ in the CHIME observing band (400--800 MHz). This detection gap coincides with the Cygnus X region -- a massive star-forming region in the Galaxy, known to contain dense ionized gas and strong degrees of turbulence \citep{pm52,wendker,schneider+06}. The alignment with known Galactic structure suggests a strong link between plasma propagation effects and the suppression of FRB detectability in the CHIME/FRB experiment. 

This study reports a detailed analysis of the detection gap, investigates the extent to which propagation effects contribute to it and identifies associated patterns in scattering and detectability. The structure of the paper is as follows. In  \S\ref{sec:obs}, we describe the observational dataset including post-detection selection criteria and classification of FRBs based on morphology. In \S\ref{sec:void-significance} we present evidence for a significant detection gap observed in the sky distribution of CHIME/FRB \cat{}. In \S\ref{sec:gapexplanation}, we explore potential observational and astrophysical explanations for the detection gap, and derive an empirical lower bound on $\tau_{\rm sc}$ responsible for it. The role of Galactic scattering in shaping FRB detectability across the full \cat{} sample is discussed in \S\ref{sec:discussion}. Our conclusions are summarized in \S\ref{sec:conclusion}.

\begin{figure*}[t!]
    \centering
    \includegraphics[width=6in]{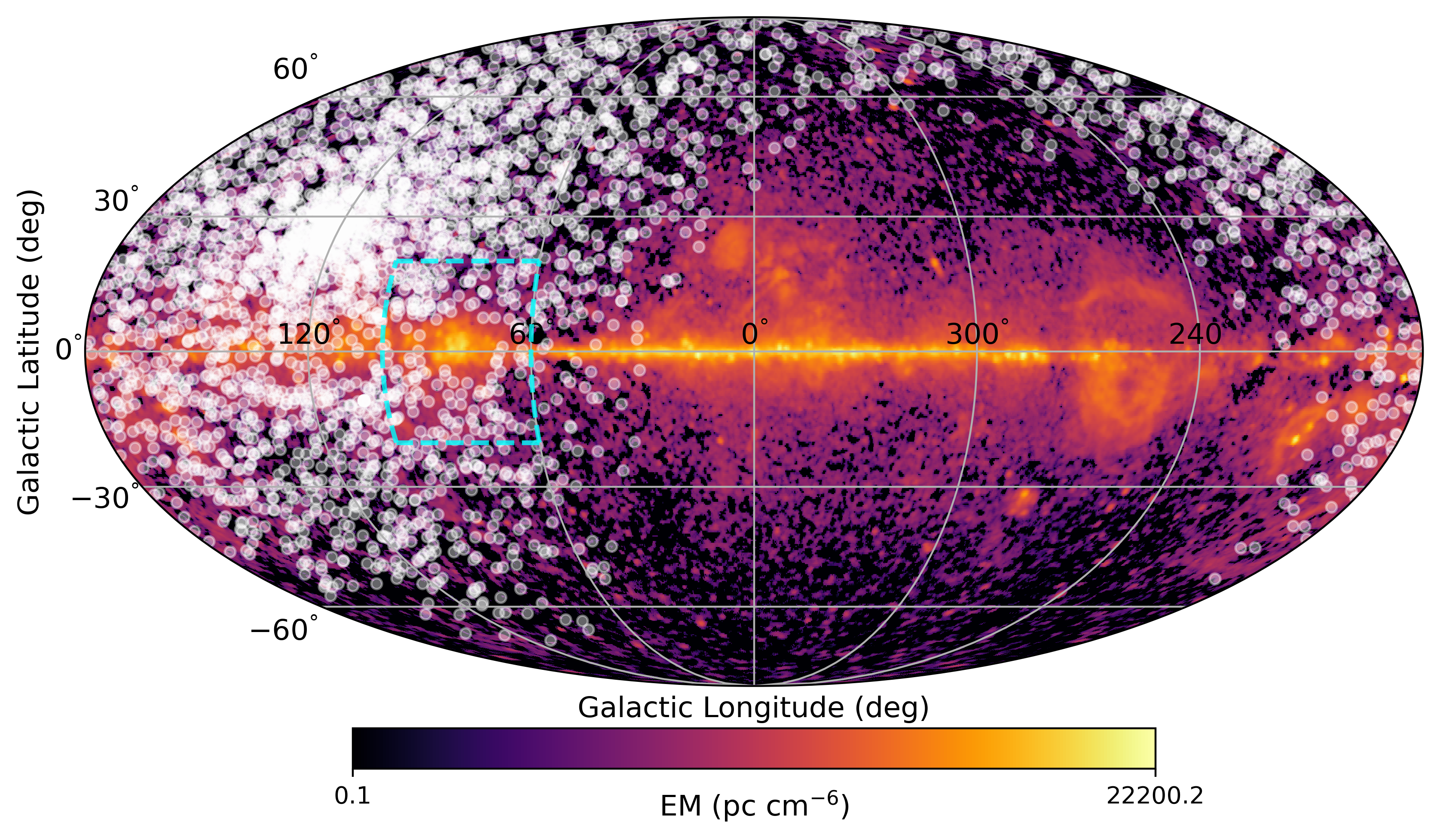}
    \caption{Positions of \numfrbs{} FRBs in our sample overlaid with the Planck emission measure map \protect{\citepalias{planckforegrounds16}}. We qualitatively see an anti-correlation between FRB detections and regions of high EM. The detection gap is  highlighted with dashed cyan lines. A zoomed-in view of this region is presented in Figure \ref{fig:literature_scattering}.}
    
    \label{fig:h-alpha-all-frb-healpy}
\end{figure*}

\section{Observations}\label{sec:obs}
In this work, we analyze data products generated by the CHIME/FRB backend for \cat{} (\citetalias{chimefrbcatalog2}). Full descriptions of the CHIME telescope and its radio-transient backends are provided in other works (\citetalias{abb+18,abb+21,abb+22}). We nonetheless summarize aspects of CHIME and the \cat{} data products relevant to our study in this section.

\subsection{The CHIME Telescope and Instrumentation}
The CHIME telescope is a compact radio interferometer that consists of 1,024 dual-polarization antennas operating in the 400--800\,MHz range. These antennae are distributed across four half-cylindrical reflectors that collectively span 80\,m$\times$100\,m. The CHIME/FRB backend receives up to 1,024 streams of beamformed, total-intensity data with time and frequency resolutions of $\rm{t_{samp}}\approx$ 0.983\,ms and $\Delta \nu_{\rm{chan}}\approx$ 24.4\,kHz, respectively \citepalias{abb+18}, and enacts a series of real-time pipelines to identify and preserve astrophysical signals for further analysis. The CHIME/FRB detection pipeline uses a cut-off of signal-to-noise ratio -- $ (S/N) \geq (S/N)_{\rm thresh} = 8 $ -- for filtering FRB candidates (\citetalias{chimefrbcatalog2}).

\subsection{FRB Celestial Positions} 
The celestial position of each FRB is computed  from total-intensity data using the procedure described by \cite{abb+19c}. This procedure uses two pieces of information -- the real-time metadata corresponding to the detecting beam,  and an analytic model of the beam shape and FRB dynamic spectrum -- to estimate a position and its statistical uncertainty. These localizations have typical uncertainties of $O(10')$.   We caution that $O({1\%})$ of bright \cat{} FRBs may be incorrectly localized to the near-sidelobes, which can cause a position offset of 0.5$\degree$--2$\degree$. This localization error does not have a significant impact on our analysis of the detection gap as the error magnitude is much smaller than the angular extent of the gap. However, this could impact the classification of H II intersections described in \S\ref{subsec:hii_intersections}. We encourage a follow-up of this work using more accurate localizations. 

\subsection{Model-dependent Classification of FRBs}
\label{subsec:model-dep-classification} 

Statistically significant signals detected by the CHIME/FRB backend are classified as ``Galactic'', ``ambiguous'', or ``extragalactic'', based on the  source position and DM estimated by the real-time  {\tt bonsai} tree de-dispersion algorithm (\citetalias{abb+18}).  Single pulses from Galactic radio pulsars are identified through coincidences in both position and DM, and for the purposes of \cat{} are excluded from subsequent analysis. Our signals of interest are those deemed extragalactic (i.e., FRBs), which are classified as such if their DMs exceed both of the predicted maximum values from the NE2001 \citep{ne2001} and YMW16 \citep{ymw17} electron density models for their lines of sight; these predictions are made assuming a distance of 25 kpc between the observer and the edge of the Milky Way disk. No aspects of scatter-broadening are considered for the real-time classification of FRBs.

\subsection{Best-fit Models of FRB Morphology} \label{subsec:morphology}
In order to characterize the effects of the WIM on burst morphology in \cat{}, we use spectro-temporal measurements of FRB morphology estimated with \fitburst{} \citep{fpb+24}. The \fitburst{} modeling framework assumes that each pulsed feature undergoes cold-plasma dispersion and, in the absence of scattering-broadening, has a Gaussian shape of intrinsic temporal width ($\sigma$). If relevant, the shape of the burst is instead assumed to be the pulse broadening function of a Gaussian profile \citep[e.g.,][]{mck14} that depends on $\sigma$ and $\tau_{\rm sc}$. 

All \cat{} \fitburst{} measurements assume that $\tau_{\rm sc}$ scales with electromagnetic frequency ($\nu$) such that $\tau_{\rm sc} \propto \nu^{-4}$. The \cat{} values for $\tau_{\rm sc}$ are originally reported referenced to $\nu=400.195$\,MHz (\citetalias{chimefrbcatalog2}); however, for this present study we scale $\tau_{\rm sc}$ to be referenced at $\nu = 1$\,GHz using $\tau_{\rm sc,\,\nu} \propto \nu^{-4}$ \citep{bcc+04}, as is typically done for related studies on radio pulsars and FRBs. Additionally, for detectability discussions, we reference  $\tau_{\rm sc}$ to CHIME frequencies whenever relevant -- explicitly denoting it as $\tau_{\rm sc,\, 600\,MHz}$. We used $\tau_{\rm sc,\,1\,GHz} > 0.13$\,ms as a threshold for determining which FRBs have ``well-measured" scatter-broadening; this threshold value was obtained through comparisons of \fitburst{} fits between total-intensity and microsecond-resolution baseband (voltage) data of the same detections \citep{sand+25_published}.

\subsection{Post-detection Selection Criteria for \cat{} FRBs}\label{subsec:criteria}
We adopted the initial data quality flags outlined in the \cat{} methodology (\citetalias{chimefrbcatalog2}). In addition, we required that the event had successful fits from the \texttt{header\_localization} and \fitburst{} pipelines. We excluded FRBs detected in the sidelobes of the primary beam, since these are more complicated to localize accurately \citep{lsn+23}. The above selection criteria resulted in 4429 FRBs. To obtain the full set of unique sightlines from \cat{}, we retain only the burst with the highest $(S/N)$ for each repeating source. The final sample utilized in this study consists of \numfrbs{} FRBs -- comprising \numnonrepfrbs{} non-repeaters and \numrepfrbs{} unique sightlines from repeating FRBs. Of those, 1105 FRBs meet the criterion for well-measured $\tau_{\rm sc,\,1\,GHz}$ described in \S\ref{subsec:morphology}.

\section{Statistical Significance of the Detection Gap}\label{sec:void-significance}
Figure \ref{fig:h-alpha-all-frb-healpy} presents the spatial distribution of FRBs in \cat{}, with the detection gap region highlighted by a cyan dashed enclosure. The high density of FRB detections near the North Celestial Pole is due to CHIME's declination-dependent sensitivity and exposure. A magnified view of the detection gap region is presented in Figure \ref{fig:literature_scattering}. To delineate its shape and extent, we applied geometric methods based on the Delaunay triangulation and its dual, the Voronoi diagram using the {\tt scipy.spatial} module \citep{scipy}.   Using the Voronoi diagram, we identified the largest empty circle within the detection gap by locating the Voronoi vertices farthest from any FRB. We estimate the center of the largest empty circle (LEC) within the detection gap to be at $(l, b)=(77.7\degree, 0.9\degree)$, with a radius of $R_{\rm LEC}=7.7\degree$. The triangulation partitions FRB positions into non-overlapping triangles such that no point is inside the circumcircle of any triangle. We identified the triangles for which the distance between their circumcenters and the center of gap fell within the radius. The convex hull obtained from connecting the outer boundary of these triangles is the polygon that encloses the sky with zero FRB detections in Figure \ref{fig:literature_scattering}, with an area of 213.6 deg$^2$. 

At 600 MHz, the effective gain of CHIME along the detection gap LOS is $G \approx1.02$ K/Jy with system temperature $T_{\rm sys}\approx$ 50 K \citep{abb+18}. Thus, the system equivalent flux density (SEFD) is roughly $T_{\rm sys}/G\approx$ 48.78 Jy. The single-pulse radiometer equation gives a minimum detectable flux density of $\sim 0.5$ Jy for narrow pulses ($\sim 1$ ms broadened width) and $\sim 0.1$ Jy for pulses with broadened width larger than 10 ms along this LOS. In the following discussion, we assess whether the apparent detection gap can be explained by instrumental sensitivity and statistical noise effects, or whether it points to suppression in detectability due to scattering caused by ionized Galactic structures. 

\begin{figure}[h]
    \centering
     \includegraphics[width=\columnwidth]{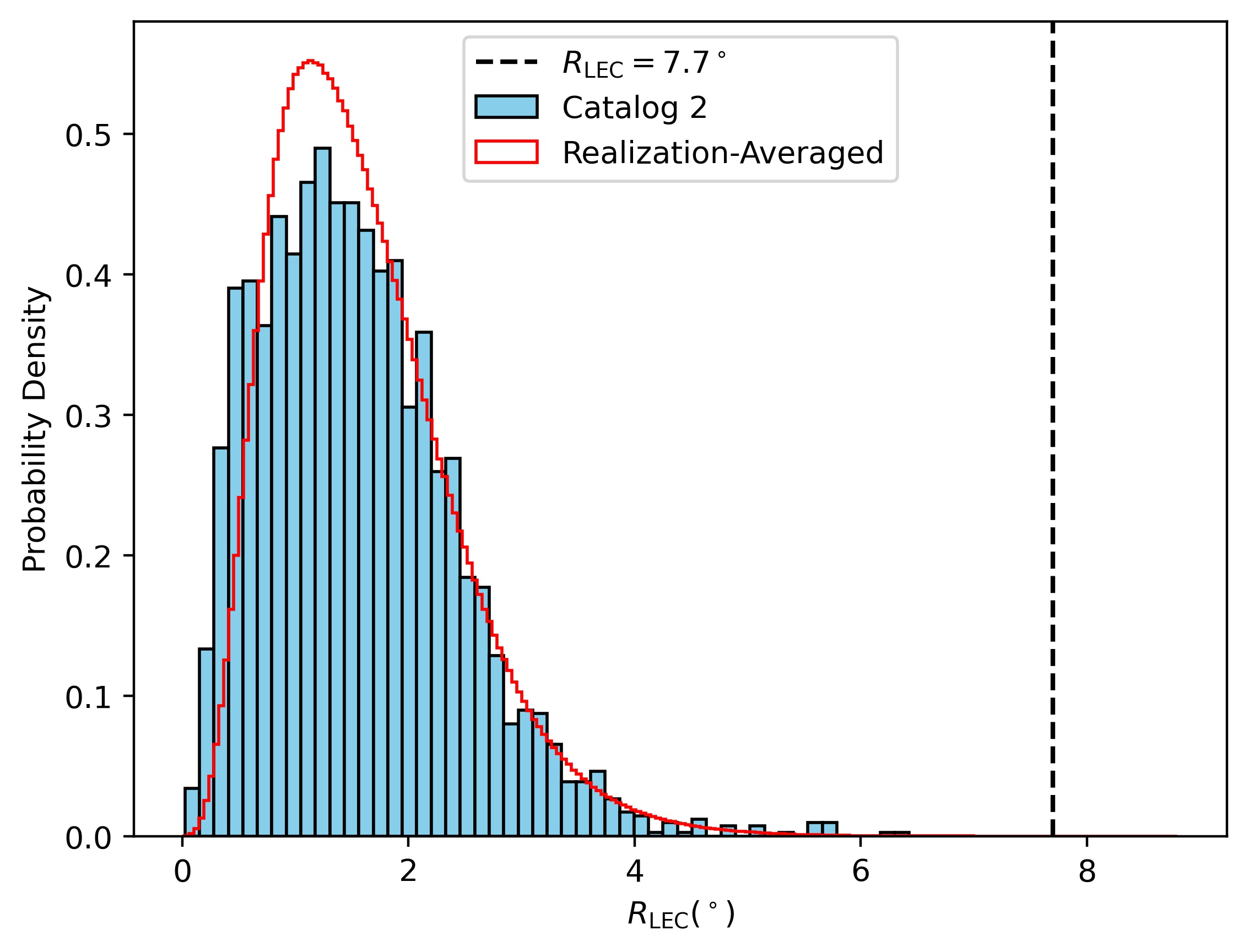}
     \caption{Distribution of $R_{\rm LEC}$ estimates obtained from a Delauney-triangulation analysis of the \cat{} data (red) and simulations of sky positions of FRBs in $N_{\rm sim} = 10^4$ catalogs derived from the statistics of \cat{} detections (blue-filled). The black-dashed vertical line indicates the largest measured $R_{\rm LEC}$ in Catalog 2.}
     \label{fig:hist_RLEC}
\end{figure}

We use two methods to evaluate the significance of the lack of detections. For the first method, we estimate the probability of obtaining detection gaps with sizes $R_{\rm LEC} \ge 7.7^\circ$ assuming that no ``foreground" process produces such gaps. We determine this probability by first using \cat{} astrometric data to derive a two-dimensional probability density distribution of FRB detections across the CHIME sky through Gaussian kernel-density estimation. We then resample this distribution to simulate a catalog of a size equal to that of \cat{}. We then execute the Delauney-triangulation algorithm described above on the subset of simulated FRBs within the range $-30^\circ < b < 30$ and $50^\circ < l < 200^\circ$ to find gaps in this realization. We repeat this process $N_{\rm sim} = 10^4$ times in order to estimate a ``realization-averaged" distribution of $R_{\rm LEC}$, which is shown in Figure \ref{fig:hist_RLEC}. Using this approach, we find that the probability of obtaining detection gaps with sizes $R_{\rm LEC} \ge 7.7^\circ$ due to statistical fluctuations in our measurements is $p \approx 1.3\times10^{-6}.$

For the second method, we compute the probability of detecting 0 FRBs within a circle of radius $R_{\rm LEC}$ or larger anywhere on the sky surveyed by CHIME/FRB assuming a non-homogeneous Poisson process model for its detections. The non-homogeneous Poisson process is described by a position-dependent intensity function $\lambda (\alpha, \delta)$ which here represents the expected density of detected FRBs and scales with the instrument's exposure and sensitivity. It is convenient to work in the equatorial coordinate system ($\alpha, \delta$) as the CHIME/FRB gain varies primarily with $\delta$. Moreover, the cosine dependence of an isotropic universal distribution of FRBs on the celestial sphere cancels out with the inverse-cosine dependence of the exposure in declination.

As reported by \cite{chimefrbcatalog1}, we note reduced exposure between $27\degree < \delta < 34\degree$ as shown in Figure \ref{fig:red-exp-dec} in Appendix \ref{sec:Appendix_exposure} and Figure 5 of \citetalias{chimefrbcatalog1}. This is due to a time-limited failure of one of the four CPU nodes covering these declinations. The patch of reduced exposure overlaps with only 7\% of the total sky area of the detection gap. Omitting the reduced exposure patch, we fit a polynomial to the logarithm of the exposure from \cat{} versus $\delta$ and find that the observed exposure in this region is $\approx18$\% lower than expected. At $\delta>70\degree$, sources are circumpolar from CHIME's latitude and cross the telescope's meridian twice daily. These `upper' and `lower' transits therefore yield higher effective exposure at $\delta>70\degree$. Using the CHIME/FRB sensitivity approximation parameterized by \citet{2024arXiv241012146C} and accounting for exposure as described above, we can write
\begin{multline}
\lambda(\alpha, \delta) = \frac{1}{A} \Big[(\cos^{1.5}(\phi - \delta))^{\gamma-1} \\ + \, H(\delta-70\degree) (\cos^{1.5} (180\degree - \phi - \delta))^{\gamma-1}\Big]
\end{multline}
where $H(x)$ is the Heaviside step function, $\gamma=1.5$ is the Euclidean power-law index for a flux-limited survey representing the FRB luminosity function, $\phi = 49.32\degree$ is the geographic latitude of the CHIME telescope (\citetalias{abb+18}), and $A$ is the following normalization constant such that the intensity equals $N_{\rm total}$, the total number of FRBs in the sample, when integrated over CHIME/FRB's sky: 
\begin{multline}
A = \frac{2\pi}{N_{\rm total}} \Bigg[
       \int\limits_{-11\degree}^{90\degree}  \left( \cos^{1.5}(\phi - \delta) \right)^{\gamma-1}d\delta\ \\
        + \int\limits_{70\degree}^{90\degree} 
        \left( \cos^{1.5}( 180\degree - \phi - \delta) \right)^{\gamma-1} d\delta \Bigg].
\end{multline}
The probability of detecting zero bursts within a circle of the observed radius $R_{\rm LEC}$ or larger anywhere on the sky can then be written as
\begin{multline}
P(N=0 \in C(R_{\rm LEC}) |\lambda)  \\ 
= \int\limits_{0\degree}^{360\degree} \int\limits_{-11\degree}^{90\degree} 
\exp \Biggl[-\hspace{-5mm}\iint\limits_{C(R_{\rm LEC}, \{\alpha_0,\delta_0\})}\hspace{-5mm} \lambda(\alpha_0, \delta_0) d\alpha_0 d\delta_0\Biggr]\cos{\delta}  d\alpha d\delta ,
\end{multline}

where $C(R_{\rm LEC}, \{\alpha, \delta\})$ denotes the circle of radius $R_{\rm LEC}$ centered at $\{\alpha, \delta\}$. This probability serves as our one-sided \( p \)-value.  We find that the non-detection of bursts in a region of this radius is statistically significant at the $4.2\sigma$ level obtained by converting $p \approx 1.3 \times 10^{-5}$ using the inverse survival function of the standard normal distribution.

\section{Assessing Possible Origins for the Detection Gap}
\label{sec:gapexplanation}

Cygnus X is one of the most massive star forming complexes in the Milky Way, rich in ionized gas.  While it is possible that the spatial gap occurs by chance from the increased sky temperature, additional plasma processes are likely at play.  Nominally, NE2001 predicts that the maximum DM through the region is $\approx 500$\dmu{}, and maximum scattering timescale in CHIME's observing band is $\tau_{sc, 600\rm{MHz}} \approx 1.4$\,ms, comparable to the time sampling. These values are too small to cause appreciable DM smearing, or reduced S/N from scattering. However, angular broadening measurements of background sources, and recent DM and scattering measurements of pulsars and a background FRB support a view in which Cygnus X is far more plasma rich than in existing Galactic electron models; in this Section we cover the existing literature, and investigate the gap in context of other observations through Cygnus X.

\begin{figure}
    \centering
    \includegraphics[width=\columnwidth]{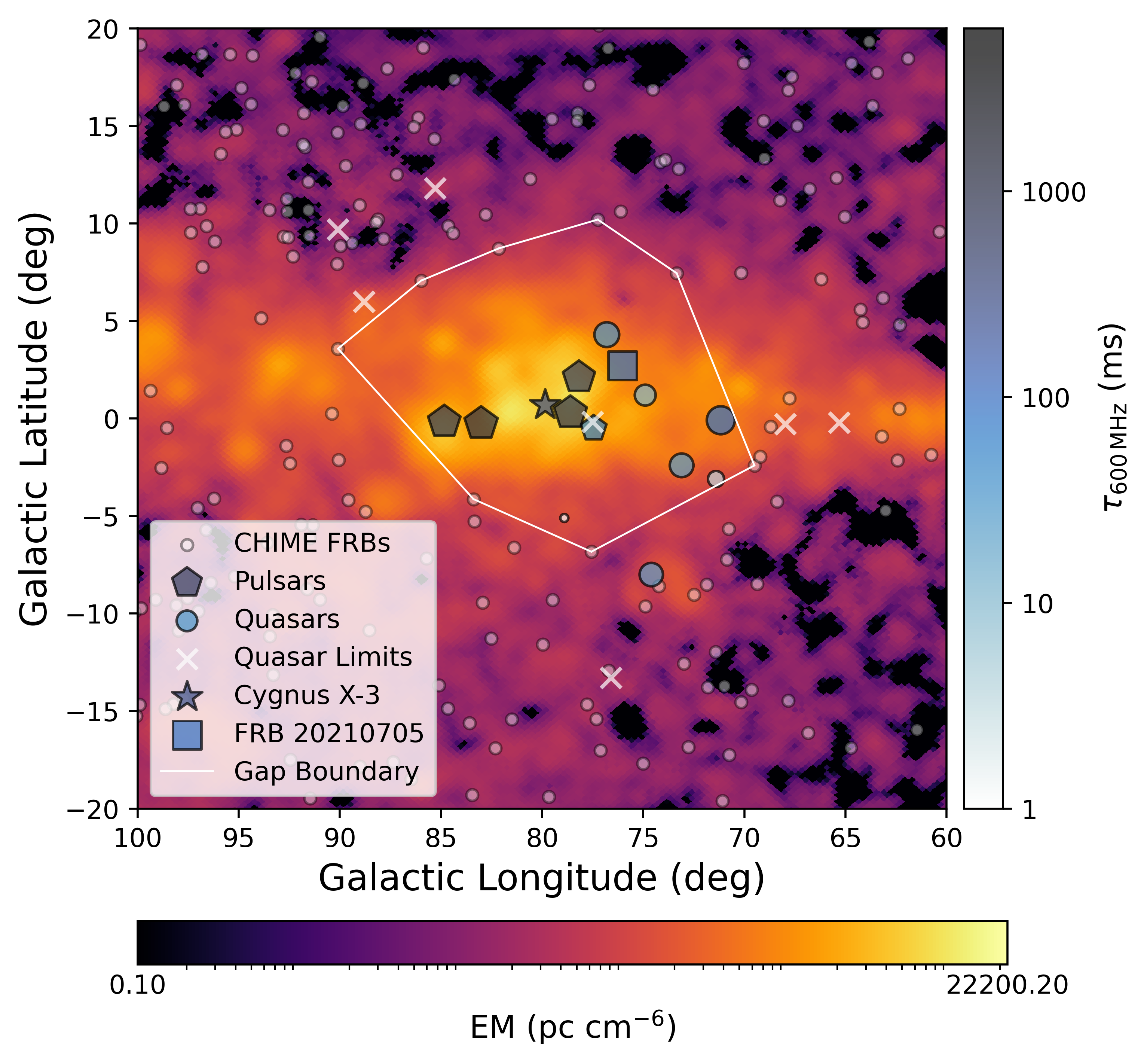}
    \caption{A comparison of literature values of scattering in and surrounding the Cygnus X region, as described in \S\ref{sec:cygnus}, overlaid on the Planck emission measure map \protect{\citepalias{planckforegrounds16}}. The angular broadening measurements for quasars and Cygnus X-3 have been converted to scattering timescales using Equation \ref{eq:tauscat_thetafwhm}, and scaled to 600\,MHz using $\theta \propto \nu^{-2}$.  Scattering measurements for pulsars and FRB 20210705 are extrapolated to 600\,MHz using $\tau \propto \nu^{-4}$. This shows that sightlines through the Cygnus X region can produce $\sim 10-1,000\,$ms of temporal scattering in CHIME's observing band, which would the hinder detection of an FRB (see \S\ref{sec:scatteringconstraint}). A white polygon encloses the area with zero FRB detections, formed by Delaunay triangulation as described in \S\ref{sec:void-significance}.}
    \label{fig:literature_scattering}
\end{figure}

\subsection{Background on the Cygnus X region}
\label{sec:cygnus}

While Cygnus X is comprised of many distinct star forming regions, studies of molecular line observations have argued that they are associated, and at a similar distance $d_{\rm Cyg}$ \citep{schneider+06}. This association was later supported using parallax measurements with masers \citep{rygl+12}, and with Gaia \citep{berlanas+19}, finding distance measurements spanning a range $1.3 < d_{\rm Cyg} <1.7$\,kpc.  We adopt a value of $d_{\rm Cyg} = 1.5$\,kpc throughout our work (following discussion  of \citealt{comeron+20}), but emphasize that it does not strongly affect our analysis. At $d_{\rm Cyg} = 1.5$\,kpc, the $\sim 18\degree$  angular diameter of the spatial gap corresponds to $\sim 250\,$ pc. Additionally, the electron temperature $T_{\rm e}$ is a function of Galactocentric radius $R_{\rm gal}$, fit  by \citet{quireza+06} as $T_{\rm e}(\text{K}) = (5780 \pm 350) + (287 \pm 46) R_{\rm Gal}.$ We adopt $T_{\rm e} = 8000\,$K throughout for the Cygnus X region, appropriate for $R_{\rm Gal} \approx 8\,$kpc.

\subsubsection{Emission Measure}
\label{sec:cygnus-em}
From the Planck emission measure (EM) map (\citetalias{planckforegrounds16}), which traces the free-free emission from ionized gas and has a resolution of $13.7' \approx 3\,$pc,  the maximum EM of the Cygnus X region is $9400$\emu. On smaller angular scales, the Cygnus X region contains many thin filamentary structures, which are overdense and overpressured compared to the average properties of their surroundings \citep{emig+22}.  The maximum EM in these filaments approaches $10^5$\emu{} with a median value of 5200\emu, and electron densities of $n_e \approx 10 - 400$\cmc, compared to the median of $n_e \approx 35$\cmc{}. The density of the volume-filling ionized gas in the region is $n_e \approx 35$\cmc{}, which itself is far above the  volume-averaged plasma density in the WIM of $n_e \approx 0.03$\cmc{}  (\citealp{mezger+78, emig+22}).

The observables $\text{DM} = \int_0^L n_e(l) dl$ and $\text{EM} = \int_0^L n_e(l)^2 dl$ are physically related quantities where $n_e(l)$ is the free electron density (in \cmc) at a distance $l$ (in pc) along the line of sight, and L is the total path length through the ionized medium (in pc). Following \citet{OAL+24}, we assume that a path length L with volume-average electron density $n_e$ is filled with dense cloudlets with a volume filling factor $f \leq 1$, and internal density $n_{e,i}$, such that $n_{e} = f n_{e,i}$.  The other parameters are the variance of electron density within a cloudlet $\epsilon^2 = \langle (\delta n_{e,i})^2\rangle / n_{e,i}^2 \leq 1$, and between cloudlets $\zeta = \langle n_{e,i}^2\rangle / \langle n_{e,i}\rangle^2 \geq 1$.  Following this prescription, the DM and EM are related as
\begin{equation}
    \text{DM} = \sqrt{\frac{f}{\zeta(1+\epsilon^2)}\ \text{EM}\times\text{L}} \ .
    \label{eq:DMEM}
\end{equation}
The prefactor $\dfrac{f}{\zeta(1+\epsilon^2)}$ is always $\leq 1$, forming an upper limit on the $\text{DM}$.

Based on the above, a source seen through Cygnus X at the maximum $\text{EM} \approx 10^4$\emu\ (from the Planck map) with path length $\rm L \approx  500\,$pc could accrue $\text{DM} \lesssim 2200$\dmu. As will be described in \S\ref{sec:cygnus-pulsars}, \S\ref{sec:cygnus-fastfrb}, this estimate is in line with the highest DMs of pulsars in the region, and that of the background \frb{} discovered at L-band with the Five-hundred-meter Aperture Spherical radio Telescope (FAST).  As discussed in \S\ref{sec:cygnus-em}, the peak EM in filaments reaches $10^5$\emu, although most filaments have thickness on the order of a few pc \citep{emig+22}. A sightline could, in principle, accumulate a larger DM if it is seen directly through the long axis of a filament, but this is unlikely, and not representative of an average line-of-sight through the Cygnus X region.

\subsubsection{Angular Broadening}
\label{sec:cygnus-theta}
Background quasars, if intrinsically compact enough, can show signs of angular broadening when observed through multi-frequency very long baseline interferometry (VLBI), characterized by broadening $\theta \propto \nu^{\alpha}$, scaling with $\alpha \approx -2$. Early VLBI observations of background quasars show angular broadening with full-width at half-maximum (FWHM) of $\theta_{\rm FWHM} \approx 10-200\,$ mas referenced to 1\,GHz within Galactic longitudes $60\degree \lesssim l \lesssim 80\degree$, greater than the surroundings \citep{fey+89, fey+91}.  Observations of Cygnus X-3 (not associated with, but background to the Cygnus X region), show scatter broadening of $220-250\,$ mas at 1\,GHz \citep{CygnusX-molnar}. Many additional VLBI studies of quasars through Cygnus X (e.g. \citealt{lazio+01, desai+01, gabanyi+06}) show angular broadening of a similar order, as well as extreme scattering event-like structures seen in the light-curve of quasar B2005+403, implying $\sim 0.7\,$AU plasma structures \citep{koryukova+23}.

Angular broadening, combined with the known distance to Cygnus X, can be used to predict the scattering time delay. For a scatter-broadened image with $\theta_{\rm FWHM}$, the scattering timescale $\tau_{\rm sc}$, defined as the $1/e$ timescale for a decaying exponential, is related to $\theta_{\rm FWHM}$ as  (e.g. \citealt{cc19}):
\begin{equation}
    \tau_{\rm sc} \approx \frac{d_{\rm eff} \theta_{\rm FWHM}^2}{8 \ln(2)c},
\label{eq:tauscat_thetafwhm}
\end{equation}
where $d_{\rm eff} \equiv \dfrac{d_{\rm scr}(D-d_{\rm scr})}{D}$, for a scattering screen at distance $d_{\rm scr}$ from the observer and source at distance $D$. For an extragalactic source and a thin screen near the observer, $d_{\rm scr}\ll(D-d_{\rm scr})$, and $d_{\rm eff} \approx d_{\rm scr}$

This equation is exact for an angular intensity distribution following a Gaussian, and can differ by a factor of $\approx 0.6-2$ depending on the exact form of the scatter-broadened image. Scattering timescales derived for background radio sources using Equation \ref{eq:tauscat_thetafwhm} with $d_{\rm scr}=d_{\rm Cyg}$, are listed in Table \ref{table:literature_t_scat}.

\subsubsection{Pulsars}
\label{sec:cygnus-pulsars}
There are $> 90$ known pulsars within our detection gap \citep{manchester+05}.  With surveys from more sensitive telescopes, many newly discovered pulsars in this region have DMs greater than the full expected Milky Way contribution from Galactic-$n_e$ models \citep{han+25}, as the abundance of plasma within Cygnus X is not properly accounted for.  Without a reliable $n_e$ model along these sightlines, and with few independent distance measurements, it is difficult to know which pulsars are foreground or background to the star-forming complex. As the complex itself is likely the cause of the discrepancy, the sources with DMs far exceeding model predictions can be assumed to be within or behind Cygnus X.

The FAST telescope recently measured scattering of 17 pulsars within our detection gap, spanning a range of $\tau_{\rm sc,\, 1\textrm{ GHz}} \approx 1-300$\,ms, and DMs $200-950$\dmu{} \citep{jing+25}.  Other pulsars with scattering measurements include PSR J2108+5001 discovered by the CHIME All-sky Multi-day Pulsar Stacking Search (\citetalias{CHAMPSS+25}) 
located just outside of our detection gap boundary, and 9 pulsars discovered with the Green Bank Telescope (GBT) at 820 MHz \citep{mcewan+24}.

The maximum DM predicted by NE2001 and YMW16 at the center of the detection gap is $\sim 500$ pc cm$^{-3}$. Pulsars with DMs exceeding this value are almost certainly located within or behind Cygnus X. Their scattering timescales can therefore provide an estimate of the propagation effects experienced by background sources, including FRBs intersecting this region. A list of the literature scattering measurements for $\text{DM} > 500$\dmu{} pulsars across the Cygnus X region is compiled in Table \ref{table:literature_t_scat} and in Figure \ref{fig:literature_scattering} (including J2108+5001 (482\dmu{}) which falls just below this threshold but as a CHIME-detected pulsar is relevant to the context of CHIME's observing capabilities).

\subsubsection{FRBs at Higher Observing Frequencies}
\label{sec:cygnus-fastfrb}
According to Blinkverse \citep{blinkverse}, a single FRB lies within the CHIME detection gap: \frb{} (see Figure \ref{fig:literature_scattering}), discovered by FAST in the Galactic Plane Pulsar Snapshot survey at L-band \citep{zhj+23}. This burst is located at Galactic coordinates (76.03$\degree$, 2.71$\degree$) with DM $= 2011.6 \pm 3.2$ pc cm$^{-3}$ and $\tau_{\rm sc,\,1\,GHz} =  12 \pm 2$ ms ($\tau_{\rm sc,\,600\,MHz} \approx 92 \pm 15$ ms).

The distribution of observed scattering timescales in \cat{} is well-described by a log-normal distribution (\citetalias{chimefrbcatalog2}). In log-space, ${\rm log}_e(\tau_{\rm sc}/\rm{ms})$ follows a normal distribution characterized by a mean $\mu = -0.5$ and standard deviation $\sigma=1.02$. This corresponds to a mean $\tau_{\rm sc,\, 1\textrm{ GHz}} = 1.02$ ms and $\sigma = 1.37$ ms. \frb{} with ${\rm log}_e(\tau_{\rm sc}/\rm{ms}) = 2.48$, lies 2.93$\sigma$ above the mean. This value corresponds to the 99.5th percentile of the \cat{} distribution, placing it among the most highly scattered FRBs in the sample.

We include all aforementioned measurements of angular broadening, and scattering timescales in Figure \ref{fig:literature_scattering}, superimposed on the Planck foreground EM map, surrounding our spatial gap in FRB detections.

\subsection{Biases from DM Models and Dispersion Smearing}\label{subsec:dmsmear}
As described in \S \ref{subsec:model-dep-classification}, the CHIME/FRB pipeline uses Galactic DM predictions from NE2001 and YMW16 to classify events as Galactic or extragalactic. This raises a general concern that DM-based thresholds could cause nearby, low-DM FRBs to be missed, if the models overestimate the Galactic DM along the LOS. For example, FRB 20220319D (DM = 110.95\dmu, distance = 50 Mpc) detected by the Deep Synoptic Array (DSA-110), exhibits a DM lower than the model predictions \citep{rcc+25}. However, in the Cygnus X region, the situation is reversed and the Galactic-$n_e$ models underestimate the DM (see pulsars in \S\ref{sec:cygnus-pulsars}). As a result, pulsars with unusually high DMs are more likely to be misclassified as FRBs, not vice versa. Therefore, this classification bias cannot explain the absence of FRBs in the detection gap.

We acknowledge the possibility that the Cygnus X complex is so poorly accounted for in the models that they could underestimate the maximum DM by an order of magnitude, i.e., that the true value of ${\rm DM}_{\rm max}\ge 5000\textrm{ pc cm}^{-3}$ along the LOS. At such high DMs, intra-channel dispersion smearing could partially emulate scatter-broadening and reduce pulse brightness at low frequencies. However, we believe that DM smearing can be ruled out as the origin of the detection gap. Temporal broadening due to intra-channel dispersion smearing \citep{cm03} is given by:
\begin{equation}
    \rm{t_{chan}} = 8.3\mu s\left(\frac{\Delta \nu_{\text{chan}}}{\text{MHz}}\right)\left(\frac{\nu}{\text{GHz}}\right)^{-3}\left(\frac{\text{DM}}{\text{pc cm}^{-3}}\right).
\label{eq:tchan}
\end{equation}
 
The CHIME/FRB detection pipeline searches to a maximum DM of 13000\dmu{} (\citetalias{abb+18}). In order to produce $\rm t_{\rm chan} \approx 8$ ms at 600 MHz -- comparable to the mean value of $\tau_{\rm sc}$ measured in \cat{} (1 ms at 1GHz; see \S\ref{sec:cygnus-fastfrb}) --  a value of ${\rm DM} > 8000 \textrm{ pc cm}^{-3}$ is required. This estimate is substantially higher than the DM inferred from the maximum EM in the Cygnus X region as discussed in \S \ref{sec:cygnus-em}.

\subsection{Free-free Absorption in the Direction of Cygnus X}
\label{subsec:free-free}
 
Here, we examine whether the FRB detection gap towards Cygnus X could be caused by free-free absorption.
The free-free opacity is well approximated by \citep{mezger67}
\begin{equation}
    \tau_{ff} = 3.28\times10^{-7} \left(\dfrac{T_e}{10^4\,\rm K}\right)^{-1.35} \left(\dfrac{\nu}{\rm GHz}\right)^{-2.1}  \left(\dfrac{\rm EM} {\text{ pc cm}^{-6}} \right),
\end{equation}

As discussed in \S\ref{sec:cygnus}, the Planck EM map for the Cygnus X region indicates that $\rm{EM} \approx 9400$\emu, while detailed studies of overdense and overpressured filaments show peak values of $\rm{EM} \approx 10^{5}$\emu{} \cite{emig+22}.

Inserting $\rm{EM} \approx 10^4$\emu\ for the Cygnus X region \citep{OAL+24}, and $T_e \approx 8000\,$K, the opacity is $\tau_{ff} \approx 0.02$ at $\nu = 400\,$MHz, the bottom of our band.  For the maximum measured emission measures of $\rm{EM} \approx 10^5$\emu, free-free absorption begins to matter, with $\tau_{ff} \approx 0.2$ at $\nu = 400\,$MHz, but still a small effect, and with $\tau_{ff} < 0.1$ averaged across the band.  The detection of background quasars at $350$\,MHz (e.g. \citealt{lazio+01}), and prevalence of free-free emission at $148\,$MHz (e.g. \citealt{emig+22}) also qualitatively suggests that free-free absorption is not dominating at low radio frequencies in this region.  We therefore conclude that while free-free absorption could have a small effect on a background FRB, it is not the primary cause of the detection gap.

\subsection{Effect of the Sky Temperature}
\label{sec:skytemperature}
The spatial gap is towards Cygnus X \citep{piddington+52}, a bright source of radio continuum at CHIME's observing frequencies which could reduce sensitivity to a background burst.
Using the Haslam 408 MHz radio sky temperature map \citep{haslam}, scaled to the central frequency of the CHIME band with a scaling index of $-2.6$, the average sky temperature within the detection gap is $35$\,K. To capture the effect of higher temperatures present in smaller sub-regions within the area, we use the 99th percentile $T_{\rm sky} =108$\,K.

The expression for $(S/N)_b$, the signal-to-noise ratio of the broadened pulse, 
\begin{equation}
    (S/N)_b = \frac{GS_i\sqrt{n_p\Delta\nu}}{\beta(T_{\rm sys}+T_{\rm sky})} \frac{W_i}{\sqrt{W_b}}
    \label{eq:snr_b}
\end{equation}
follows directly from the standard radiometer equation as used in FRB and pulsar literature (e.g., \citealp{cm03, pab+18}),
where $S_i$ is the intrinsic flux density, $n_p$ is the number of polarizations measured, $\Delta\nu$ is the bandwidth and $\beta$ is the correction factor for digitization. Burst detectability for CHIME/FRB is governed by the intrinsic burst width $W_i$, intrinsic flux, DM and $\tau_{\rm sc}$. The observed peak flux density for an FRB decreases as its broadened pulse width - $W_b$ - increases, defined as \citep{gcl+21} 
\begin{equation}
    W_b  = \sqrt{W_i^2(1+z_0)^2 + \tau_{\rm sc}^2 + t_{\rm samp}^2 + t_{\rm chan}^2},
    \label{eq:broadened_width}
\end{equation}
where $z_0$ is the redshift and $t_{\rm chan}$ depends linearly on the DM as described by Equation \ref{eq:tchan} in \S\ref{subsec:dmsmear}. 

To compute the expected number of bursts within the detection gap, we perform simulations using subsamples of bursts from \cat{}, explicitly accounting for the effect of elevated sky temperature. To ensure approximately uniform exposure and sensitivity, we select all \cat{} bursts in the similar declination range of $30\degree < \delta < 50\degree$, resulting in 919 bursts. This sample serves as an empirical proxy for the intrinsic burst sample seen at this declination range in the absence of the influence of Cygnus X. From \S\ref{sec:void-significance}, the expected number of bursts originating behind the gap (i.e., before accounting for the elevated $T_{\rm sky}$ in the Cygnus X region) is 19, following a Poisson distribution. We generate 10$^6$ simulated realizations as follows:

\begin{itemize}
    \item Generate a number of bursts originating behind the gap $N_{\rm gap}$, drawn from a Poisson distribution $P(\lambda=19)$;
    \item Randomly select $N_{\rm gap}$ bursts from the \cat{} subsample of 919 bursts, using their intrinsic burst width ($W_i$), sky temperature corresponding to their positions $T_{\rm sky, i}$ and signal-to-noise  $(S/N)_i$ as burst parameters;
    \item Compute the $(S/N)$ resulting from the changed sky temperature in the gap, as: 
    \begin{equation*}
       (S/N)_{\rm Cygnus} = (S/N)_{i} \dfrac{T_{\rm sky, i} + T_{\rm sys}}{T_{\rm sky, Cygnus} + T_{\rm sys}} ;
    \end{equation*}
    \item For each realization, find the number of bursts detected in the gap, $N_{\rm detected,gap}$, such that $(S/N)_{\rm Cygnus}\geq8$.
\end{itemize}

From the simulations, we find the mean number of bursts detected would be $\langle N_{\rm detected,gap}\rangle = 6$, when accounting for the above-average $T_{\rm sky}$.  We find that zero detections occur in fewer than $0.3\%$ of the simulations (i.e., $N_{\rm detected,gap}=0$), indicating that increased sky temperature and Poisson fluctuations alone are insufficient to account for the observed gap towards Cygnus X.

\subsection{Empirical Limits on Scattering Timescale}
\label{sec:scatteringconstraint}
As summarized in \S\ref{sec:cygnus}, the measurements to date of sources observed through the Cygnus X region -- through angular broadening of background quasars and Cygnus X-3, scattering of distant pulsars and one FRB detected by FAST -- show significant scattering ($\tau_{\rm sc,\,600\,MHz}\sim10-1000$\,ms). This already suggests that scattering can be deleterious for FRB detections at CHIME, and is likely a significant contributor to the lack of CHIME/FRB detections in the gap.  However, the magnitude of scattering, which depends on plasma fluctuations on tiny scales at $\ll \rm{AU}$, can drastically differ between sightlines. In this Section, we quantify the effect of scattering on CHIME/FRB detections, and in turn set a lower limit on the average scattering in the detection gap which would lead to zero detections. 

We first searched the positional and \fitburst{} data in \cat{} for a radial trend in measured $\tau_{\rm sc}$ from the boundary of the Cygnus X region, observing no clear trend. In Catalog 1 injections, CHIME/FRB found a strong selection bias against temporally broad bursts ($W_b > 10$ ms, calculated at 600 MHz; \citealt{mts+23}). The lack of a trend may be due to the narrow observable window in $\tau_{\rm sc}$: the lower limit set by the reliability cut-off (see \S\ref{subsec:morphology}; 0.13 ms at 1 GHz); and the upper limit ($\geq 10$ ms at 1 GHz) that corresponds to scattered pulses likely too broad to detect.  Along with a limited number of detected FRBs, a statistically significant radial trend is difficult to establish without finer time resolution to probe a wider range in $\tau_{\rm sc}$. As a result, it is difficult to trace if there is a gradual increase in scattering timescales for sightlines closer to the center of the detection gap.
 
From the simulations described in \S\ref{sec:skytemperature}, we found that on an average 6 bursts are expected to be seen in the detection gap. These simulations account for the elevated sky temperature but not for scattering. Here, we extend the simulation to assess the degree of scattering required to see zero bursts. 

\begin{figure}[h]
    \centering
     \includegraphics[width=\columnwidth]{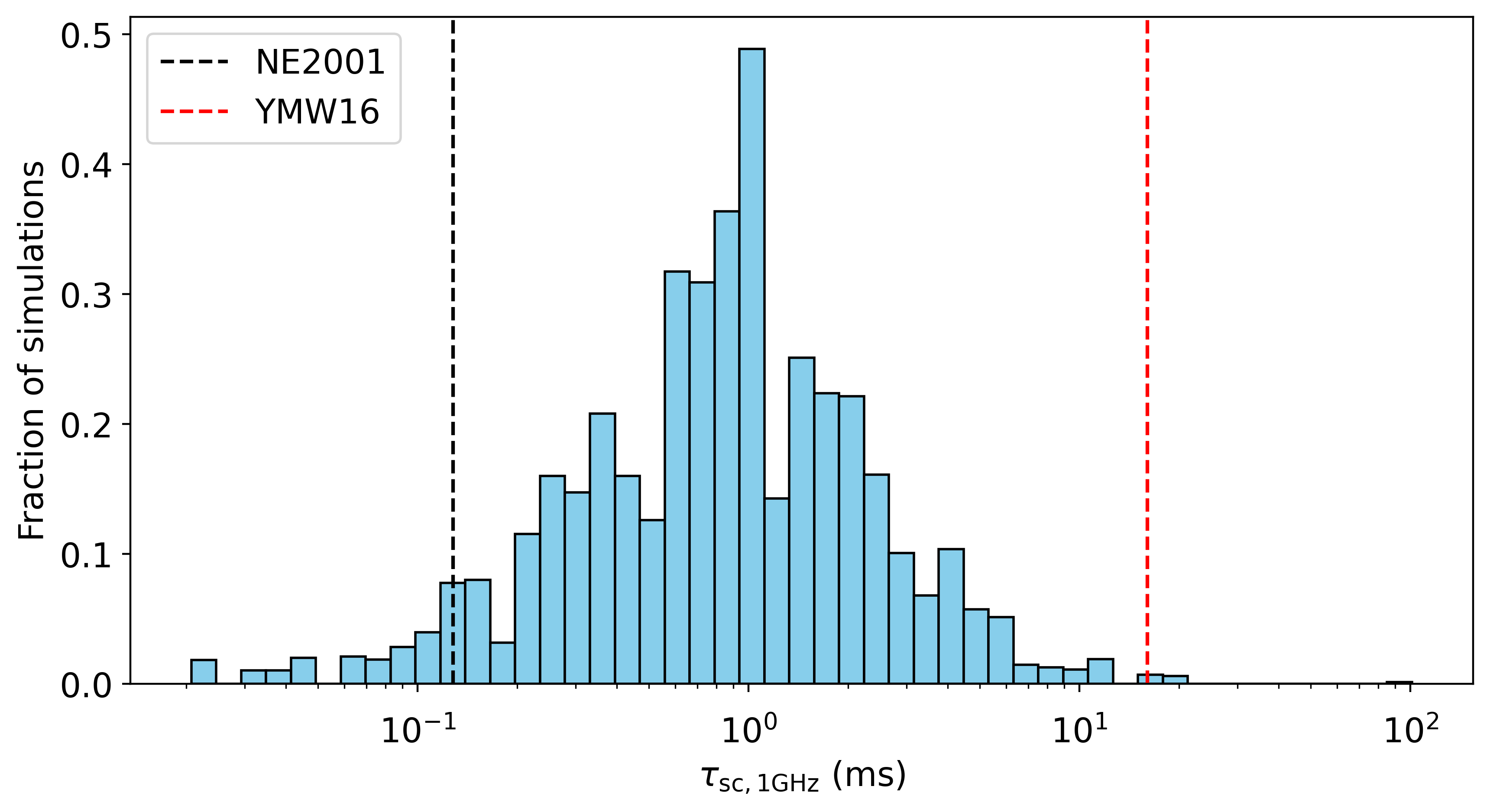}
     \caption{Distribution of scattering timescales  at 1 GHz that result in zero detectable FRBs in the gap across 10$^6$ simulations, incorporating elevated sky temperature as described in \S\ref{sec:skytemperature}. Vertical dashed lines indicate scattering predictions from the NE2001 model (black) and YMW16 model (red) for the line of sight along the center of the gap.}
     \label{fig:t_scat_lim}
\end{figure}

In the simulations outlined in \S\ref{sec:skytemperature}, we introduce pulse broadening due to scattering using Equation \ref{eq:broadened_width}. For each simulated realization, we compute the minimum $\tau_{\rm sc}$ using Equation \ref{eq:snr_b}, such that the increased $W_b$ and $T_{\rm sky}$ reduce the signal-to-noise of all FRBs within the detection gap to $(S/N)_b < (S/N)_{\rm thresh} = 8$. The distribution of $\tau_{\rm sc,\, 1GHz}$ values obtained from these simulations is shown in Figure \ref{fig:t_scat_lim}. We find that a mean value of $\tau_{\rm sc,\,1GHz} = 5.59$\,ms, or $\tau_{\rm sc,\, 600MHz} = 43.13$\,ms is sufficient to suppress all detections. Because all simulations adopt the 99th percentile $T_{\rm sky}$ measured within the void area, these are conservative lower limits on $\tau_{\rm sc}$. Using more typical, lower $T_{\rm sky}$ would require even larger $\tau_{\rm sc}$ to achieve the same suppression in $(S/N)_b$. The maximum $\tau_{\rm sc}$ predicted by Galactic-$n_e$ models for the LOS through the center of the detection gap are: $\tau_{\rm sc,\, 1GHz}^{\rm NE2001}$ = 0.1 ms and $\tau_{\rm sc,\, 1GHz}^{\rm YMW16}$ = 16 ms. This range of model-dependent maximum values of $\tau_{\rm sc}$ is consistent with our inferred maximum derived from the simulations. Moreover, this consistency reinforces the idea that enhanced scattering in Cygnus X is likely to explain the detection gap, though our limit was obtained in a manner independent of either $n_e$ model. However, we emphasize that these models are outdated, differ in their treatment of the Cygnus X region\footnote{For example, the YMW16 model does not incorporate external information available on the Cyngus X complex.}, and lack accurate modeling of small-scale plasma fluctuations, so their predictions should be interpreted only as rough references.

\section{Discussion} \label{sec:discussion}

\begin{figure*}[ht!]
    \centering
    \begin{tabular}{c}
    \includegraphics[width=\textwidth]{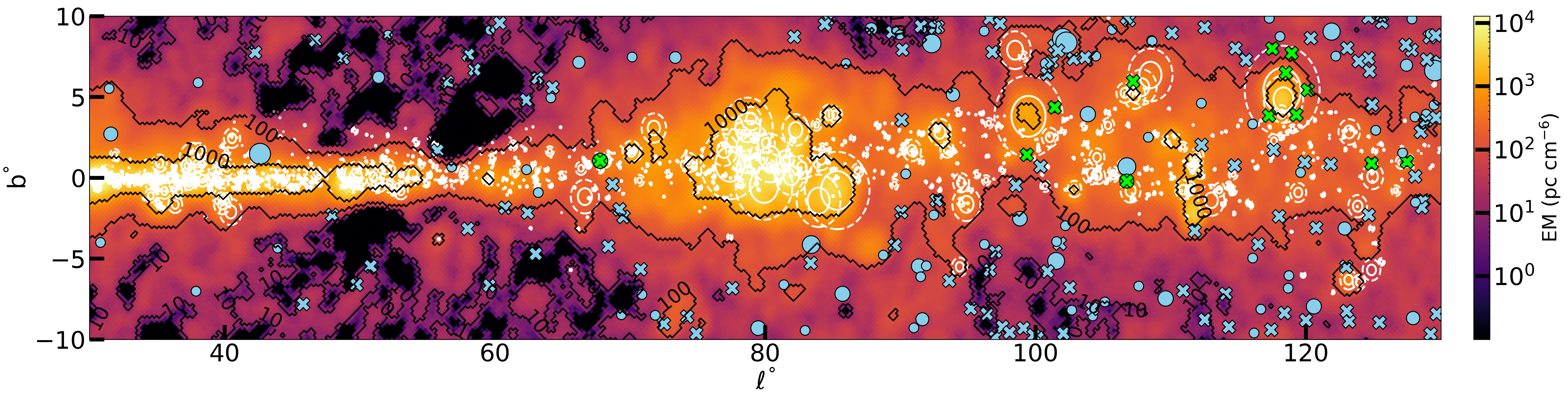}\\
    \includegraphics[width=\textwidth]{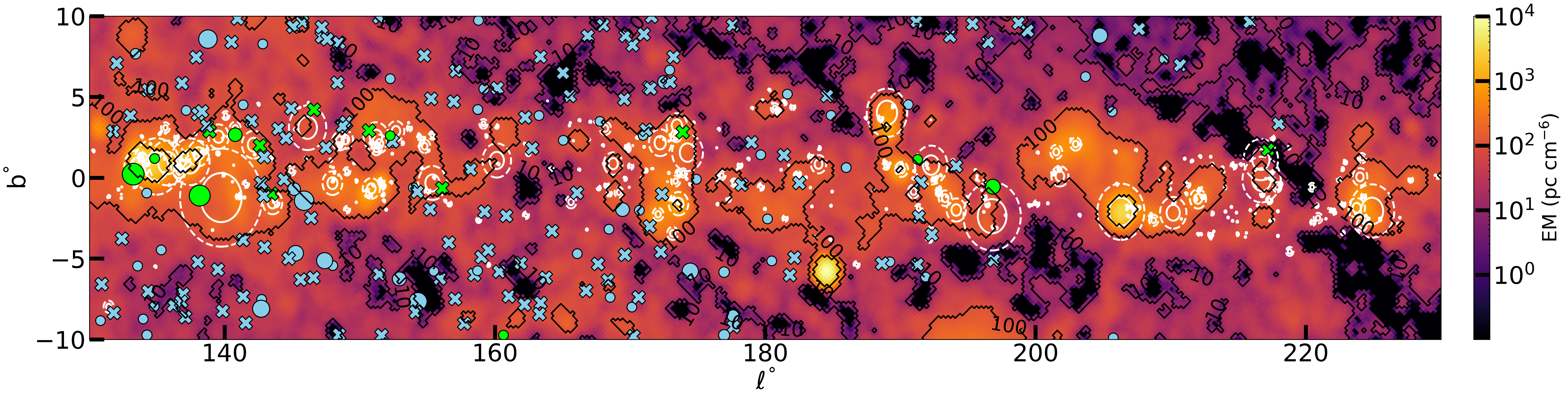}\\
    \end{tabular}
    \caption{Planck EM map (\citetalias{planckforegrounds16}) in Galactic coordinates centered at $l = 80\degree$, $b = 0\degree$ (top) and $l = 180\degree$, $b = 0\degree$ (bottom). Overlaid black contours trace logarithmically-spaced EM 10, 100 and 1,000 \emu. White solid and dashed circles highlight WISE H II  regions \citep{wise} with angular radii $\theta_\text{IR}$ and $2\theta_\text{IR}$, respectively. The radius of the IR emission $\theta_\text{IR}$ is provided by the WISE Catalog. Lime markers are for FRB LOS with intersecting H II  regions and skyblue markers are for non-intersecting FRBs. Cross markers indicate FRBs for which no reliable scattering timescale was measured with total-intensity data, while the radii of circular markers are scaled according to the log of the measured scattering timescale.}
    \label{fig:hii_grid}
\end{figure*}

 The statistical significance of the detection gap was aided by its large angular extent, as discussed in \S\ref{sec:void-significance}. Smaller spatial gaps however, can occur through chance. We used the non-detections of FRBs in the direction of Cygnus X to place a data-driven lower limit on the magnitude of maximal scattering contribution from the Milky Way Galaxy, in a manner independent of existing Galactic-$n_e$ models. However, even if smaller in angular extents, other ionized structures in the Galaxy such as H II regions and  high-EM sightlines  may have similar impacts on the observed scattering properties and detectability of FRBs in \cat{}.  This may introduce a selection bias in the observed spatial distribution of FRBs. 

The Murchison Widefield Array Interplanetary Scintillation survey has revealed similar spatial gaps in the distribution of compact radio sources due to angular broadening from clumpy ISM turbulence correlated with \halpha{} emission \citep{morgan+22}. In the case of pulsars, scintillation arcs and VLBI have been used to localize scattering structures to known H II regions (e.g. \citealt{mall+22}).

Figure \ref{fig:hii_grid} provides a visual overview of the spatial relationship between \cat{} FRBs,  known H II regions, and EM. Several FRB sightlines in our sample intersect known H II regions.  Furthermore, Figure \ref{fig:hii_grid} illustrates the possibility that FRBs with significant scattering are more likely to occur along lines of sight intersecting H II regions   -- or be missed entirely due to reduced detectability.  These visual trends in the sky distribution of \cat{} FRBs motivates the discussion in the following subsections.

\subsection{Trends with DM, EM, and \halpha{}}
\label{subsec:trends}

\begin{figure}
    \centering
    \includegraphics[width=\columnwidth]{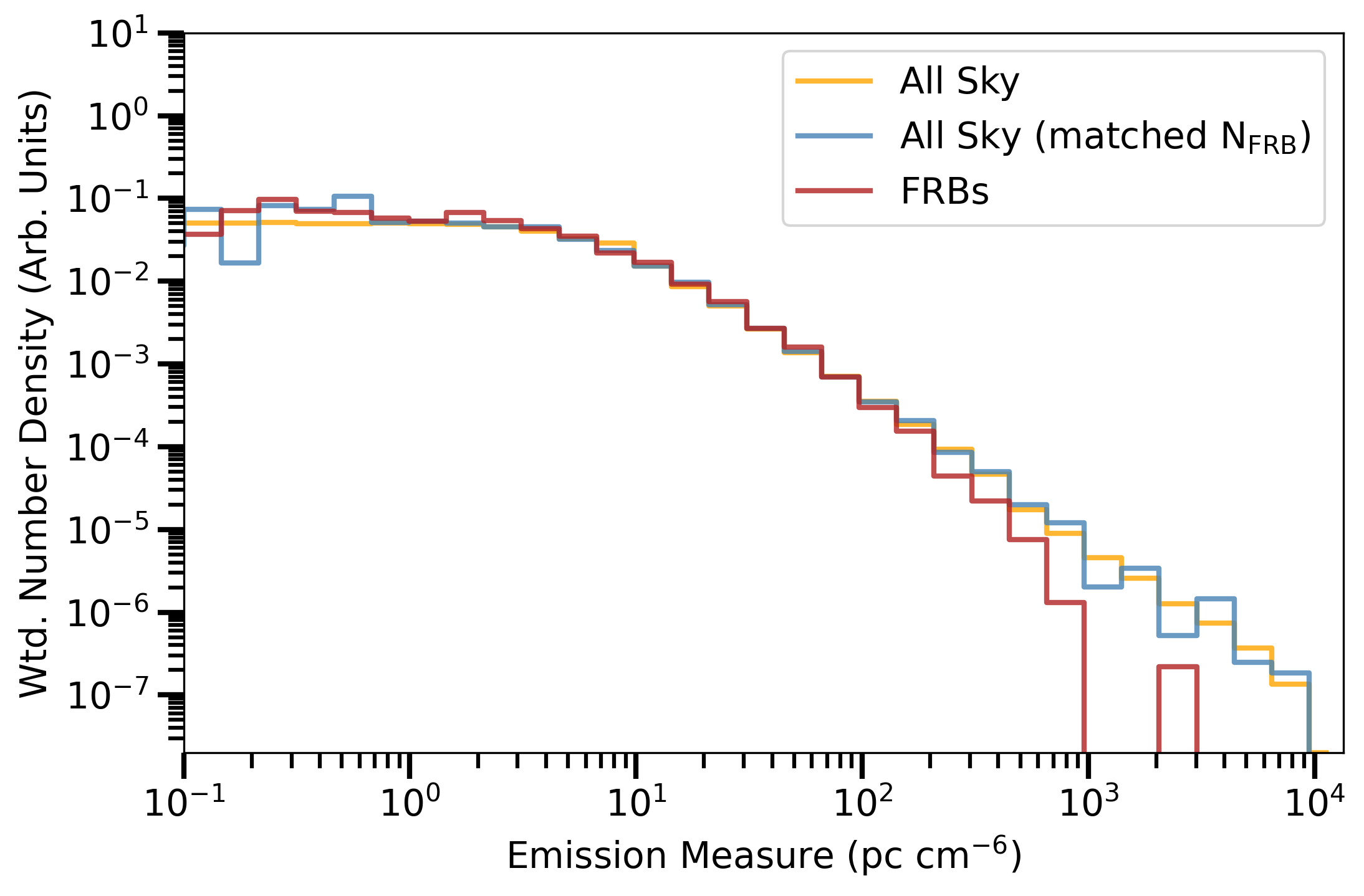}
    \caption{ Normalized, exposure weighted (wtd.) number density histogram of emission measures along all CHIME/FRB sightlines (orange) compared to the EM distribution at detected FRB positions (red). The `All Sky (matched N$_{\rm FRB}$)' (blue) curve corresponds to EMs along the same number of sightlines as the FRB sample, uniformly distributed in the CHIME-visible sky. This control sample is plotted to assess whether the apparent deficit of FRBs along high-EM sightlines arises from small-number statistics rather than a physical suppression. The y-axis is plotted in arbitrary (arb.) units for ease of comparison between samples. }
    \label{fig:EM}
\end{figure}

To statistically test whether highly scattered FRBs are under-detected in general due to propagation through regions of elevated emission measure, we compared EMs at FRB positions with those along all sightlines visible to CHIME. Exposure weighting is applied using the CHIME/FRB beam exposure map (\citetalias{chimefrbcatalog2}), such that each sky position contributes proportionally to its observing time. Figure \ref{fig:EM} shows this comparison using the Planck EM map (\citetalias{planckforegrounds16}). The FRB distribution drops to zero for EM $\geq$ 2900 pc cm$^{-6}$. This  difference demonstrates a deficit of FRBs along high EM sightlines that are associated with ionized regions in the Galaxy.  

We can ask what the above EM cutoff of 2900\emu{} corresponds to in terms of $\tau_{ff}$, DM, and $\tau_{\rm sc}$.  Similar to the calculations in \S\ref{subsec:free-free}, $\tau_{ff}$ is negligible ($\approx0.003$) in our band. Using Equation \ref{eq:DMEM}, and assuming L =  10\,pc (a typical H II region path length; see \S\ref{subsec:hii_intersections}), we estimate  a DM contribution near the cutoff of  $\approx 170$\dmu, which is too small to induce appreciable DM smearing. We estimate $\tau_{\rm sc}$ using the empirical $\tau_{\rm sc}-{\rm DM}$ relation from \citet{cordes2016radio}, resulting in a rough lower bound of $\tau_{\rm sc,\,1GHz} \gtrsim 0.5\,$ms, or $\tau_{\rm{sc,\,600MHz}}\gtrsim 4\,$ms. We note that use of their relation while assuming DM $\propto \sqrt{\rm EM}$ yields $\tau_{\rm sc} \propto {\rm EM}^{2.25}$ for DM $\gtrsim 100$\dmu; values of EM above the aforementioned cutoff yield  values of $\tau_{\rm sc}$ that rapidly rises to magnitudes that make FRBs undetectable at CHIME. We also note that pulsars behind H II regions lie preferentially above the $\tau_{\rm sc}-{\rm DM}$ relation and  that a future reassessment is therefore warranted \citep{OAL+24}.  While we caution that the logic mapping EM to $\tau_{\rm sc}$ is indirect,  the above results support the interpretation that FRBs are preferentially scatter-broadened beyond detectability along high EM sightlines.

\begin{figure}[h]
    \centering
    \includegraphics[width=\linewidth]{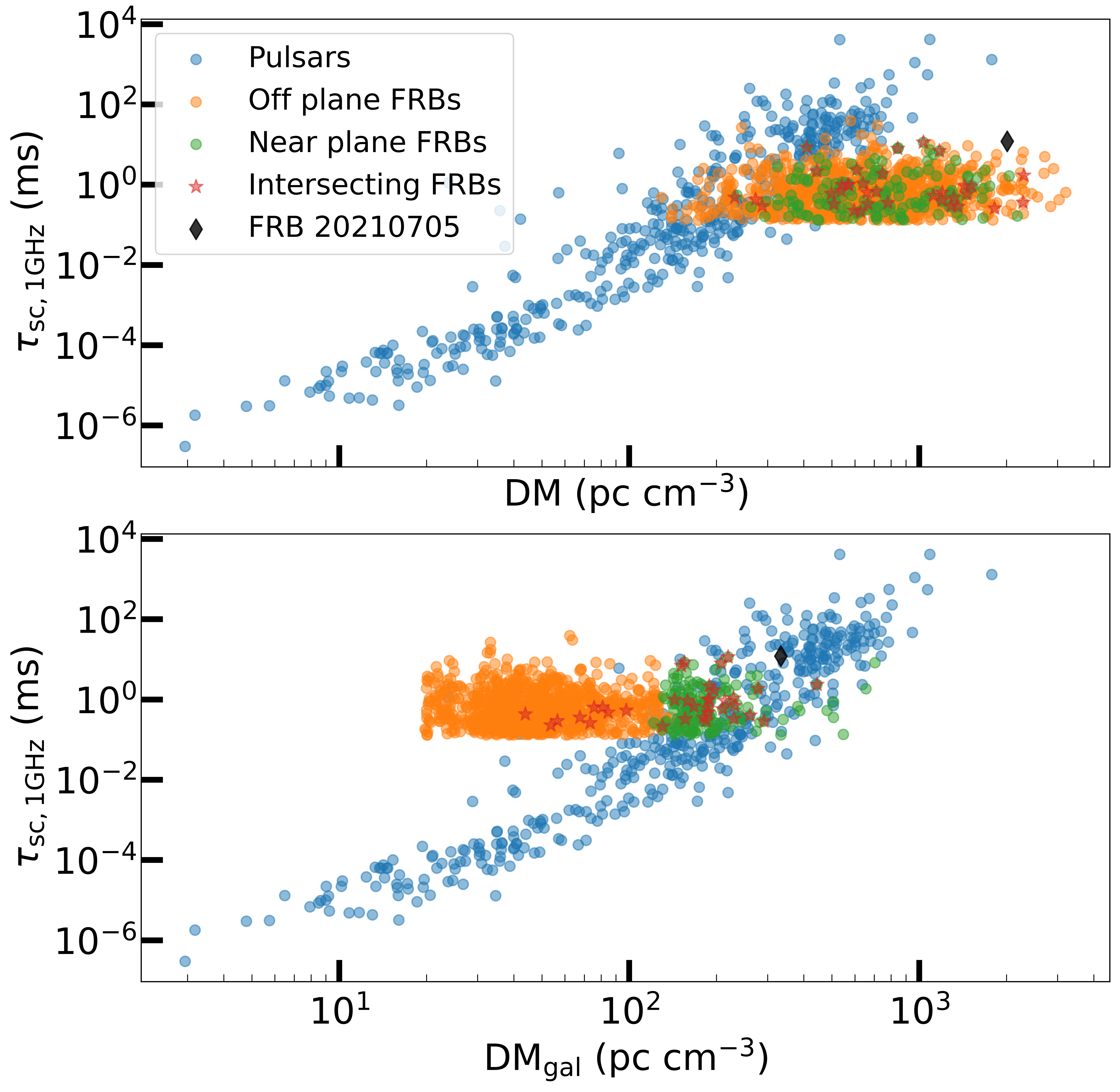}
    \caption{\cat{} FRBs plotted alongside pulsars and \frb{} in the $\tau$-DM plane.  Top panel: $\tau_{\rm sc,\, 1\,GHz}$ versus total DM. Bottom panel: $\tau_{\rm sc,\, 1\,GHz}$ versus Galactic DM contribution. In both panels, blue points show Galactic pulsars, green points show FRBs near the Galactic plane $(|b|<10\degree)$ and orange points mark those away from the plane. Red star markers show FRBs intersecting known H II regions and black diamond marks the \frb{}.}
    
    \label{fig:tau-dm}
\end{figure}

\subsection{H II Region Intersections and Scattering Properties}
\label{subsec:hii_intersections}
Similar to the treatment for pulsar sightlines intersecting H II regions by \cite{OAL+24}, we utilize the WISE V2.4 catalog of H II regions \citep{wise}. The WISE catalog contains over 8,400 H II  regions and H II  region candidates, identified using the all-sky Wide-field Infrared Survey Explorer (WISE) satellite data. We match FRBs to WISE H II regions by computing angular separations in Galactic coordinates and checking whether each FRB falls within twice the angular radius of any H II  region ($< 2\theta_{\rm IR}$). We adopt this maximum impact parameter of $2\theta_{\rm IR}$ following the prescription of \cite{OAL+24} to account for potential underestimation of the ionized gas extent by the infrared size, which may miss diffuse outer emission caused by radiation leakage \citep{luisi+19, dey+24}. In Appendix \ref{sec:Appendix_intersecting}, we report the angular separations ($\eta$) between H II regions and intersecting FRB sightline candidates, in units of $\theta_{\rm IR}$. Out of the \numfrbs{} FRBs in our sample, 36 are candidate sightlines that pass through the WISE H II  regions. We similarly examined the optical H II region catalog HH14 \citep{vizierhh14} following the method described in \cite{OAL+24}. We do not include the results in our main analysis due to the lack of angular radii in HH14, however the results are consistent with those obtained using WISE.

Among candidate FRB sightlines that intersect H II regions, as well as those that do not but lie near the Galactic plane ($|b|< 10\degree$), only about one-third pass the $\tau_{\rm sc}$ reliability cut off described in \S\ref{subsec:morphology}. This sample of sightlines is not sufficiently large to enable a robust statistical comparison of $\tau_{\rm sc}$ between these subsets. For completeness, we report the mean and median $\tau_{\rm sc,\,1GHz}$ values: for intersecting sightlines, the mean is 1.65 ms and the median is 0.63 ms; for non-intersecting near-plane sightlines, the mean is 1.06 ms and the median is 0.51 ms. The intersecting sample exhibits slightly higher values consistent with expectations, but an  AD-test between the two populations yields a statistic of 1.78 and a $p$-value of 0.06 indicating no significance. A more robust assessment will require either a larger sample from future CHIME/FRB catalogs or the use of raw baseband data with its higher time resolution to permit measurement of shorter scattering timescales. For interested readers, Table \ref{table:intersecting} in Appendix \ref{sec:Appendix_intersecting} lists the FRB sightlines that intersect H II regions, along with their positions, DMs, $\tau_{\rm sc}$, and angular separations in units of $\theta_{\rm IR}$. These may be useful for further case studies of scattering in dense Galactic environments. Their dynamic spectra are presented in Figure \ref{fig:waterfall} in Appendix \ref{sec:Appendix_intersecting} to illustrate observed pulse morphologies.

For future investigation, we point out two interesting candidates. The FRB 20190912D LOS passes through the Sharpless 171 (S171) H II  region surrounding the   Berkeley 59 cluster of young stars. The FRB 20200416A LOS passes through the giant W4 complex enclosing the Heart Nebula, hosting numerous O/B-type stars and the IC1805 open cluster of young stars. Interestingly, the \fitburst-measured $\tau_{\rm sc}$ for FRB 20200416A is smaller than the maximum predicted by NE2001 for an extragalactic source along this LOS. This could indicate an overestimate by the model, or geometric suppression of scattering if the dominant screen lies at a cosmological distance/closer to the observer. However, we note that NE2001 was not designed to predict scattering with high precision, especially for extragalactic sources and along complex or poorly constrained sightlines. As such, this comparison should be interpreted cautiously. 

We also explore the possibility of a correlation or lack thereof between $\tau_{\rm sc}$ and DM for FRBs. Figure \ref{fig:tau-dm} shows a $\tau_{\rm sc}$-DM plot for all FRB candidates in \cat{}, and exhibits an overall trend that confirms previous findings -- FRBs are systematically under-scattered compared to expectations from the $\tau_{\rm sc}$-DM relation from Galactic pulsars \citep{cordes2016radio}. The discrepancy arises because a large fraction of the FRB DM comes from the intergalactic medium, which contributes minimally to scattering \citep[][]{lbm+07, tsb+13}. However, this comparison is  biased to reflect the statistics of detected events  as we likely miss highly scattered FRBs passing through H II regions   that therefore fall below the  CHIME/FRB detection threshold.  For example, \cat{} FRBs that intersect Galactic H II regions do not show a strong correlation between DM and $\tau_{\rm sc}$ in Figure \ref{fig:tau-dm} for FRBs -- even when considering only the Galactic contribution to DM -- despite the expectation that significant $\delta n_e$ in H II regions causes scattering. However, \frb{} as detected by the FAST telescope (see discussion in \S\ref{sec:cygnus-fastfrb}) indeed follows the $\tau_{\rm sc}-{\rm DM}$ trend, indicating that its scattering is consistent with being induced by the Cygnus X region. While the statistics currently remain small, this difference raises a cautionary note for population studies: low scattering, high DM bursts may be overrepresented in \cat{}. CHIME/FRB's selection function for Catalog 1 has been studied by injecting synthetic FRBs \citep{mts+23}. This effort confirmed that the CHIME/FRB backend is more sensitive to bursts with smaller broadened pulse widths. An updated injections study for \cat{} is currently underway. In order to make conclusions about the cosmological population of FRBs, a follow-up study incorporating the updated selection functions and host models is essential.

\subsection{Visibility Fitting of Scattered FRBs as a Complementary Probe of ISM Scattering}

As discussed in \ref{sec:cygnus-theta}, a scattered source will also be angularly broadened, where larger time delays correspond to larger deflection angles.  Assuming $d_{\rm eff} \approx d_{\rm scr}$ for an extragalactic source, 
the mapping between delay $\tau$ and observed angle $\theta$ for each scattered path through a thin screen is (e.g. \citealt{wmsz04})
\begin{equation}
    \theta = \sqrt{\frac{2c\tau}{d_{\rm scr}}} \approx 91\text{ mas} \left(\frac{\tau}{10\,\text{ms}}\right)^{1/2} \left(\frac{d_{\rm scr}}{1\,\text{kpc}}\right)^{-1/2}.
\end{equation}

The CHIME/FRB Outriggers project \citepalias{aaa+25_published} augments CHIME/FRB by providing milliarcsecond-scale localization using VLBI across three strategically placed stations in North America. By enabling precise FRB localization, the Outriggers facilitate host galaxy identification, studies of FRB environments, and their use as cosmological probes. Given CHIME/FRB Outriggers' target resolution of $\theta_{\rm res}\approx 50\,$ mas \citepalias{aaa+25_published}, FRBs scattered by $\tau_{\rm sc,\,600\,MHz}\gtrsim 10$\,ms at $\sim$kpc distances will   have an angular extent larger than the Outriggers' resolution. Many of the bright bursts piercing through H II  regions, or one detected through Cygnus X (with $\tau_{\rm sc,\,600\,MHz} \gtrsim 40$ ms), would be easily resolved by CHIME/FRB Outriggers. We note that while CHIME/FRB is biased against finding bursts of these scattering times, particularly bright bursts are seen with sufficient scattering for this approach (e.g. see \citealt{shin+24} for a bright burst scattered $\gtrsim 1\,$s in the CHIME band.)

Given the mapping between $\tau$ and $\theta$, the interferometric visibility ($V$) of scattered FRBs (or pulsars) will show a characteristic shape as a function of $V(\tau, \nu)$ through a scattering tail \citep{wucknitz12, Wucknitz_GC}, which allows for a determination of screen distances and orientations.   Conversely, for spatially unresolved sources, an upper limit on the angular broadening will place a lower limit on the screen distance. This will allow mapping of Galactic scattering structures to astrophysical sources, and will cleanly distinguish between Galactic and extragalactic scattering screens.

 Upcoming ultra-wideband instruments such as the Canadian Hydrogen Observatory and Radio Transient Detector \citep[CHORD; 300--1500 MHz;][]{vlg+19} and DSA-2000 \citep[0.7--2GHz;][]{hrw+19} will be particularly well-suited to studies of scattering, offering improved sensitivity across a broad frequency range.

\section{Conclusion}\label{sec:conclusion}

Our analysis shows that FRB detections in CHIME/FRB \cat{} exhibit a two-dimensional dependence on Galactic coordinates. The most striking feature is the complete absence of CHIME/FRB detections in a patch of the sky that coincides with the Cygnus X region. The analyses discussed in  \S\ref{sec:void-significance}--\ref{sec:discussion} demonstrate that this detection gap, along with other regions of reduced FRB detections within the Galactic plane, corresponds with known H II regions. Based on the discussion presented in \S\ref{sec:discussion}, we favor the interpretation that turbulence along these lines of sight induces large degrees of scatter-broadening that totally smear FRBs out of detectability within the CHIME band. 

As shown in Section \ref{sec:gapexplanation}, measurements of $\tau_{\rm sc}$ in CHIME/FRB \cat{} and derived $\tau_{\rm sc}$ from non-detections can be used to gauge the accuracy of existing $n_e$ models of the Milky Way Galaxy. We specifically used the CHIME/FRB \cat{} measurements of $\tau_{\rm sc}$ to argue that the lower limit on scattering in the direction of Cygnus X is $\tau_{\rm sc,1GHz} \ge 5.59$ ms in order to produce zero detections in the CHIME band. This estimate is inconsistent with the NE2001 prediction for the maximum $\tau_{\rm sc}$; and while consistent with the YMW16 prediction, this latter model does not explicitly account for individual H II regions. Our analysis of FRB $\tau_{\rm sc}$ measurements therefore demonstrates that FRB measurements can be used to place ``background" constraints on inhomogeneous $n_e$ models in a manner that complements the ``foreground" constraints placed by radio pulsars. This prospect of using FRB $\tau_{\rm sc}$ measurements as tools to constrain $n_e$ models is especially important in an era where cosmological applications of FRBs are often sensitive to knowledge and/or reasonable assumptions of the $n_e$ distribution for both the Milky Way and FRB-host galaxies \citep[e.g.,][]{zkz25}.

Future CHIME/FRB measurements of $\tau_{\rm sc}$ will increase the surface density of FRB detections at low frequencies. Such an increase will likely yield additional detection gaps associated with Galactic H II regions that have smaller angular extents. These same data sets will also allow for data-driven constraints on the Galactic electron density models themselves  , through all-sky estimates of the maximum $\tau_{\rm sc}$ as estimated for the Cygnus X region in \S \ref{sec:scatteringconstraint}. Although the nature of FRB classification makes similar constraints on DM more complicated, phenomenological models can be used to gauge the likely maximum values for the contribution of the Milky Way to DM \citep[e.g.,][]{cbg+23}. We therefore expect that future CHIME/FRB catalogs will provide the means to strengthen and expand upon the inferences made in our work.

The CHIME/FRB experiment is also recording baseband (voltage) data for the high-significance subset of its catalogs \citep{mmm+21}. Access to baseband data allows for evaluation of polarization profiles \citep[e.g.,][]{mgm+23, mgm+23b, mbe+25, ppm+24}, scattering-broadening at the sub-ms level \citep{sand+25_published, curtin+25_published}, and the presence of scintillation down to the Nyquist-limited data resolution \citep[e.g.,][]{npb2024}. A catalog of such  FRB measurements will provide novel avenues for exploring the structure of the WIM , e.g., using FRB scintillation as a tool to complement the analysis of intra-day scintillation of blazars coincident with Galactic H$\alpha$ emission \citep[e.g.,][]{kjh+19}. These measurements may also be useful for exploring the impact of other Galactic features that interact with the WIM \citep[e.g., the ``Fermi bubbles;"][]{ssf10,kbh20}. Furthermore, future CHIME/FRB baseband catalogs will shed light on the role of scintillation in potentially affecting FRB detection rates across Galactic latitudes \citep{mj15}. 

The CHIME/FRB experiment is also expanding its capabilities to measure the angular broadening of FRBs through Outriggers.  Coupled with the time delay from scattering, this will enable the determination of the geometry of scattering screens. This additional constraint will be crucial for further probing the role of the WIM in modulating FRB visibility. This includes spatial correlations with known plasma structures such as H II regions, \halpha{} filaments, and detailed comparisons to emission and dispersion measures across the sky. This present work can therefore be viewed as a step towards using FRBs, both in the time and image domains, as ``backlights" to explore the WIM of the Milky Way Galaxy.

\section*{ACKNOWLEDGEMENTS}
\allacks
\defcitealias{CHAMPSS+25}{The CHAMPSS Collaboration et~al.~(2025)}
\clearpage
\appendix
\section{Literature Scattering Values near Cygnus X}
\begin{table}[h]
\caption{ Table with literature values of scattering and angular broadening towards Cygnus X, from background pulsars, FRBs, quasars, and Cygnus X-3.
References: [1] \citet{jing+25}, [2] \citetalias{CHAMPSS+25}, [3] \citet{zhz_2022}, [4] \citet{fey+89}, [5] \citet{fey+91}, [6] \citet{CygnusX-molnar}
}
\centering
\begin{tabular}{lccccccc}
\hline
Source & $\ell$ (deg) & $b$ (deg) & $\theta$ (mas) & $\tau_{\mathrm{1\,GHz}}$ (ms) & $\tau_{\mathrm{600\,MHz}}$ (ms) & DM (\dmu) & Reference \\
\hline
PSR J2021+4024g & 78.18 & 2.11 & -- & 8.21e+01 & 6.33e+02 & 678.8 & [1] \\
PSR J2030+3818g & 77.45 & -0.51 & -- & 7.80  & 6.02e+01 & 596.8 & [1] \\
PSR J2030+3944g & 78.60 & 0.30 & -- & 1.14e+02 & 8.80e+02 & 934.8 & [1] \\
PSR J2046+4253g & 83.02 & -0.26 & -- & 1.27e+02 & 9.80e+02 & 622.2 & [1] \\
PSR J2052+4421g & 84.84 & -0.17 & -- & 9.50e+01 & 7.33e+02 & 543.8 & [1] \\
PSR J2108+5001 & 91.20 & 1.47 & -- & 1.91 & 14.7 & 482.3 & [2] \\
FRB 20210705 & 76.03 & 2.71 & -- & 1.19e+01 & 9.20e+01 & 2011.6 & [3] \\
1923+210 & 55.60 & 2.30 & <16.30 & <1.74e-01 & <1.34  & -- & [4] \\
1954+513 & 85.30 & 11.80 & <4.30 & <1.21e-02 & <9.34e-02 & -- & [4] \\
2005+403 & 76.80 & 4.30 & 79.50 & 4.14  & 3.19e+01 & -- & [4] \\
2013+370 & 74.90 & 1.20 & 46.40 & 1.41  & 1.09e+01 & -- & [4] \\
2021+317 & 71.40 & -3.10 & 25.70 & 4.32e-01 & 3.34  & -- & [4] \\
2022+542 & 90.10 & 9.70 & <1.40 & <1.28e-03 & <9.90e-03 & -- & [4] \\
2023+336 & 73.10 & -2.40 & 67.60 & 2.99  & 2.31e+01 & -- & [4] \\
2048+313 & 74.60 & -8.00 & 64.70 & 2.74  & 2.11e+01 & -- & [4] \\
2050+364 & 78.90 & -5.10 & 9.30 & 5.66e-02 & 4.37e-01 & -- & [4] \\
2113+293 & 76.60 & -13.30 & <1.20 & <9.42e-04 & <7.27e-03 & -- & [4] \\
3C418 & 88.80 & 6.00 & <8.00 & <4.19e-02 & <3.23e-01 & -- & [5] \\
1922+155 & 50.62 & -0.03 & <20.00 & <2.62e-01 & <2.02  & -- & [5] \\
1932+204 & 56.08 & 0.10 & <14.00 & <1.28e-01 & <9.90e-01 & -- & [5] \\
1934+207 & 56.55 & -0.07 & <9.60 & <6.03e-02 & <4.65e-01 & -- & [5] \\
1954+282 & 65.31 & -0.21 & <8.30 & <4.51e-02 & <3.48e-01 & -- & [5] \\
2001+304 & 67.97 & -0.29 & <13.60 & <1.21e-01 & <9.34e-01 & -- & [5] \\
2008+33D & 71.16 & -0.09 & 130.00 & 1.11e+01 & 8.53e+01 & -- & [5] \\
2027+383 & 77.50 & -0.17 & <189.00 & <2.34e+01 & <1.80e+02 & -- & [5] \\
Cygnus X-3 & 79.85 & 0.70 & 240 & 3.77e+01 & 2.91e+02 & -- & [6] \\
\hline
\label{table:literature_t_scat}
\end{tabular}
\end{table}

\newpage
\section{FRBs intersecting H II  regions}\label{sec:Appendix_intersecting}

\subsection{Catalog}
\begin{table}[!h]
\caption{A full list of FRB sightlines intersecting H II regions from the WISE \citep{wise} catalog.\label{table:intersecting}}
\centering
\begin{tabular}{cccc|ccccc}
\hline
TNS Name     & (l, b)    & $\tau_{\rm sc}$ & DM             & Name & (l, b) & D     & $\theta_{\rm IR}$ & $\eta$ \\ \hline
             & (deg)     & (ms at 1 GHz)   & (pc cm$^{-3}$) &      & (deg)  & (kpc) & (deg) & ($\theta_{\rm IR}$) \\ 
\hline
FRB20200306C & 15.6, 28 & -- & 337.2 & -- & 6.4, 22.9 & -- & 5.19 & 1.89\\
FRB20230511B & 15.7, 23.1 & -- & 191.6 & -- & 6.4, 22.9 & -- & 5.19 & 1.65\\
FRB20181119B & 67.8, 1 & 2.307 ± 0.043 & 608.5 & -- & 68.2, 1.1 & -- & 0.22 & 1.8 \\
FRB20230426F & 99.4, 1.4 & 0.286 ± 0.017 & 1360.1 & S131; IC1396 & 99.5, 3.8 & 0.91 ± 0.57 & 1.26 & 1.89\\
FRB20210810B & 101.4, 4.4 & 1.101 ± 0.153 & 639.6 & S131; IC1396 & 99.5, 3.8 & 0.91 ± 0.57 & 1.26 & 1.58\\
FRB20210103A & 106.7, -0.2 & 1.881 ± 0.185 & 746.2 & S142 & 107, -0.8 & 3.52 ± 0.47 & 0.37 & 1.83 \\
FRB20230102D & 107.2, 6 & -- & 983.0 & S141 & 107.9, 5.6 & -- & 0.55 & 1.37 \\
 &  &  &  & S150 & 108.5, 6.4 & 1.34 ± 0.47 & 0.82 & 1.63 \\
FRB20230316H & 117.3, 3.9 & -- & 496.1 & S171 & 118.2, 5.4 & -- & 1.39 & 1.32 \\
FRB20220411A & 117.5, 8 & 0.323 ± 0.059 & 507.4 & S171 & 118.2, 5.4 & -- & 1.39 & 1.95 \\
FRB20210309F & 118.5, 6.5 & -- & 547 & S171 & 118.2, 5.4 & -- & 1.39 & 0.82 \\
FRB20200828D & 118.9, 7.7 & 0.96 ± 0.052 & 1516.6 & S171 & 118.2, 5.4 & -- & 1.39 & 1.74 \\
FRB20190912D & 119.3, 3.9 & -- & 786.2 & S171 & 118.3, 4.9 & 0.84 ± 0.31 & 0.73 & 1.32 \\
 &  &  &  & S171 & 118.2, 5.4 & -- & 1.39 & 1.35 \\
FRB20220218A & 120.1, 5.4 & -- & 594.5 & S171 & 118.2, 5.4 & -- & 1.39 & 1.85 \\
FRB20220330A & 124.9, 0.9 & 0.367 ± 0.023 & 664.9 & -- & 125, 0.9 & -- & 0.12 & 0.95\\
FRB20220814A & 127.5, 1 & 0.919 ± 0.104 & 1430.7 & -- & 127.6, 1.1 & -- & 0.2 & 0.59\\
FRB20190302A & 133.2, 0.2 & 11.28 ± 0.604 & 1033.8 & S190; W3 & 133.5, 0.8 & -- & 0.46 & 1.31\\
FRB20200416A & 134.8, 1.2 & 0.702 ± 0.035 & 538.6 & W4; S190 & 135.8, 0.9 & -- & 0.69 & 0.6\\
 &  &  &   & W4; S190 & 135, 0.7 & 3.31 ± 0.35 & 0.87 & 1.46\\
FRB20200708A & 138.2, -1.1 & 7.998 ± 0.667 & 842.5 & -- & 139.7, -1.2 & -- & 1.52 & 1.99\\
FRB20191209B & 138.9, 2.9 & -- & 910.7 & -- & 139.6, 2.5 & 3.81 ± 0.34 & 0.38 & 1.05\\
FRB20221129E & 140.8, 2.7 & 1.709 ± 0.372 & 2293.1 & -- & 140.8, 3.1 & -- & 0.21 & 1.9\\
FRB20210207E & 142.6, 2 & -- & 479 & -- & 142.2, 2 & -- & 0.43 & 1.08 \\
FRB20211001B & 143.5, -1 & -- & 378.7 & S203 & 143.6, -1.5 & 3.39 ± 0.25 & 0.36 & 1.57\\
FRB20230709B & 146.6, 4.2 & -- & 514.9 & -- & 146.1, 3.1 & -- & 0.69 & 1.75\\
FRB20200717B & 150.7, 2.9 & -- & 295.2 & S207 & 151.2, 2.5 & -- & 0.46 & 1.46\\
FRB20230418E & 152.3, 2.6 & 0.592 ± 0.051 & 497.5 & S210 & 152.7, 2.9 & 1.67 ± 0.34 & 0.31 & 1.65 \\
FRB20200112E & 156.1, -0.6 & -- & 913.7 & -- & 155.4, -0.3 & -- & 0.53 & 1.49\\
FRB20211231B & 158.6, -11.9 & -- & 905.7 & California Nebula; S220 & 160, -12.7 & 0.77 ± 0.31 & 1.58 & 1.12 \\
FRB20211120B & 159,-18.7 & -- & 359.1 & -- & 159.9, -18.6 & 0.24 ± 0.02 & 0.78 & 0.96\\
FRB20221012E & 159.1, -15.1 & 0.457 ± 0.08 & 1118.3 & California Nebula; S220 & 160, -12.7 & 0.77 ± 0.31 & 1.58 & 1.57 \\
FRB20201017C & 159.7,-11.3 & 0.659 ± 0.048 & 731.9 & California Nebula; S220 & 160, -12.7 & 0.77 ± 0.31 & 1.58  & 0.91\\
FRB20200323G & 160.6, -9.7 & 0.212 ± 0.05 & 627.9 & California Nebula; S220 & 160, -12.7 & 0.77 ± 0.31 & 1.58  & 0.95 \\
FRB20190902B & 173.9, 2.8 & -- & 758.2 & S235 & 173.5, 3.2 & -- & 0.38 &1.54 \\
FRB20190103B & 191.3, 1.1 & 1.008 ± 0.121 & 540.7 & -- & 192.3, 0.8 & -- & 0.6 & 1.78\\
FRB20230905D & 196.8, -0.5 & 2.163 ± 0.006 & 443.7 & S268 & 196.8, -2.4 & -- & 1.06 & 1.84\\
FRB20191014B & 196.9,-12.1 & -- & 278.2 & S264; lambda Ori & 195.3, -12.1 & -- & 0.84 & 1.73 \\
FRB20211204B & 217.2, 1.8 & -- & 425.9 & -- & 216.7, 1.1 & -- & 0.64 & 1.28 \\
\hline
\end{tabular}
\end{table}

\newpage
\subsection{Waterfall Plots}

\begin{figure*}[h!]
\centering
    \includegraphics[width=\textwidth]{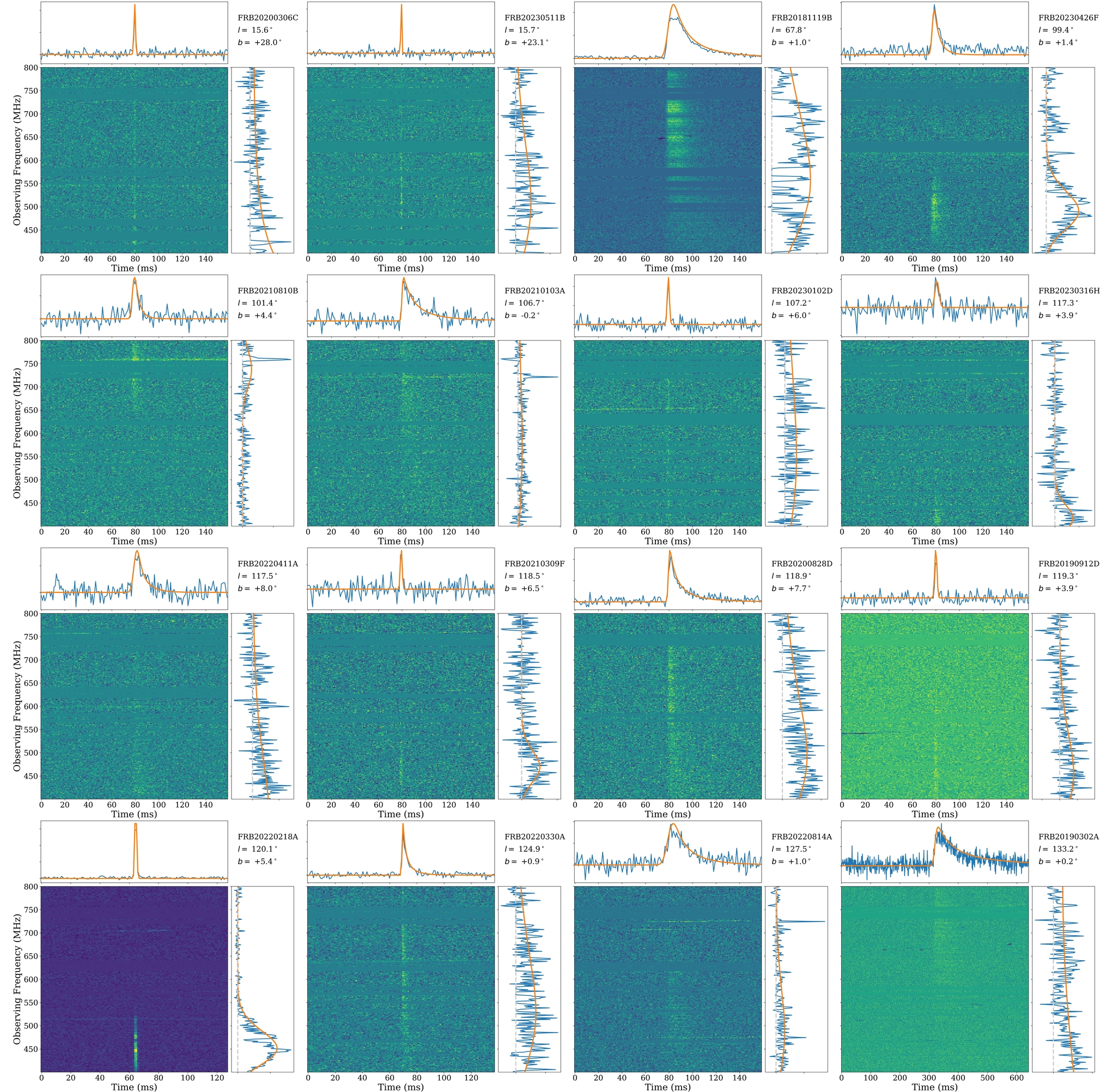}
    \caption{Observed dynamic spectra (colorized maps), time-averaged spectra (right-hand panels), and band-averaged timeseries (top panels) for CHIME/FRB detections that intersect known H II regions as described in \S\ref{subsec:hii_intersections}. The orange lines in the one-dimensional panels represent the best-fit shapes and spectral energy distributions determined by \fitburst.}
    \label{fig:waterfall}
\end{figure*}

\begin{figure*}[h!]
    \figurenum{8}
    \centering
    \includegraphics[width=\textwidth]{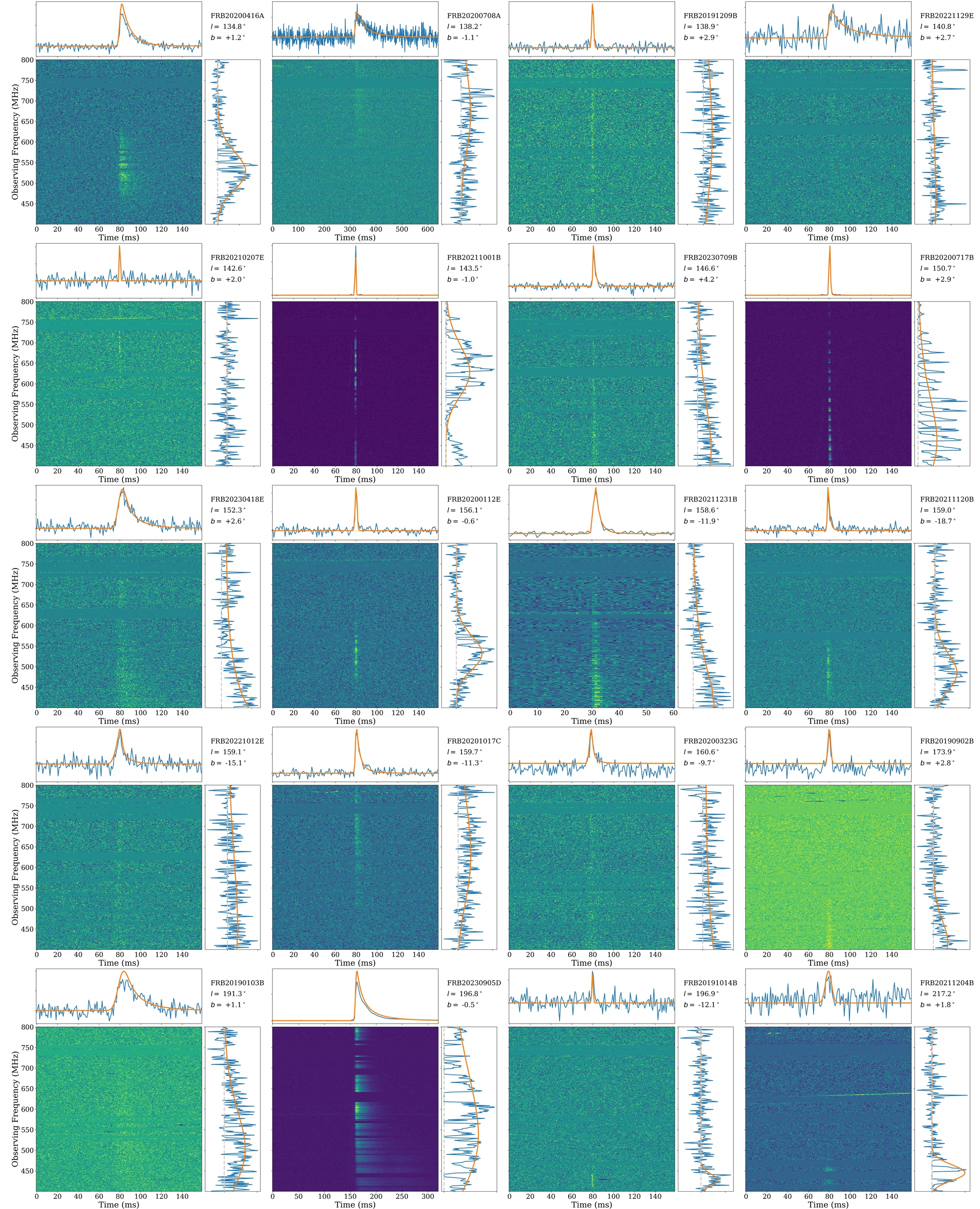}
    \caption{Continued.}
\end{figure*}

\clearpage
\section{Reduced Exposure} \label{sec:Appendix_exposure}
\begin{figure}[!h]
    \figurenum{9}
    \centering
    \includegraphics[width=0.7\columnwidth]{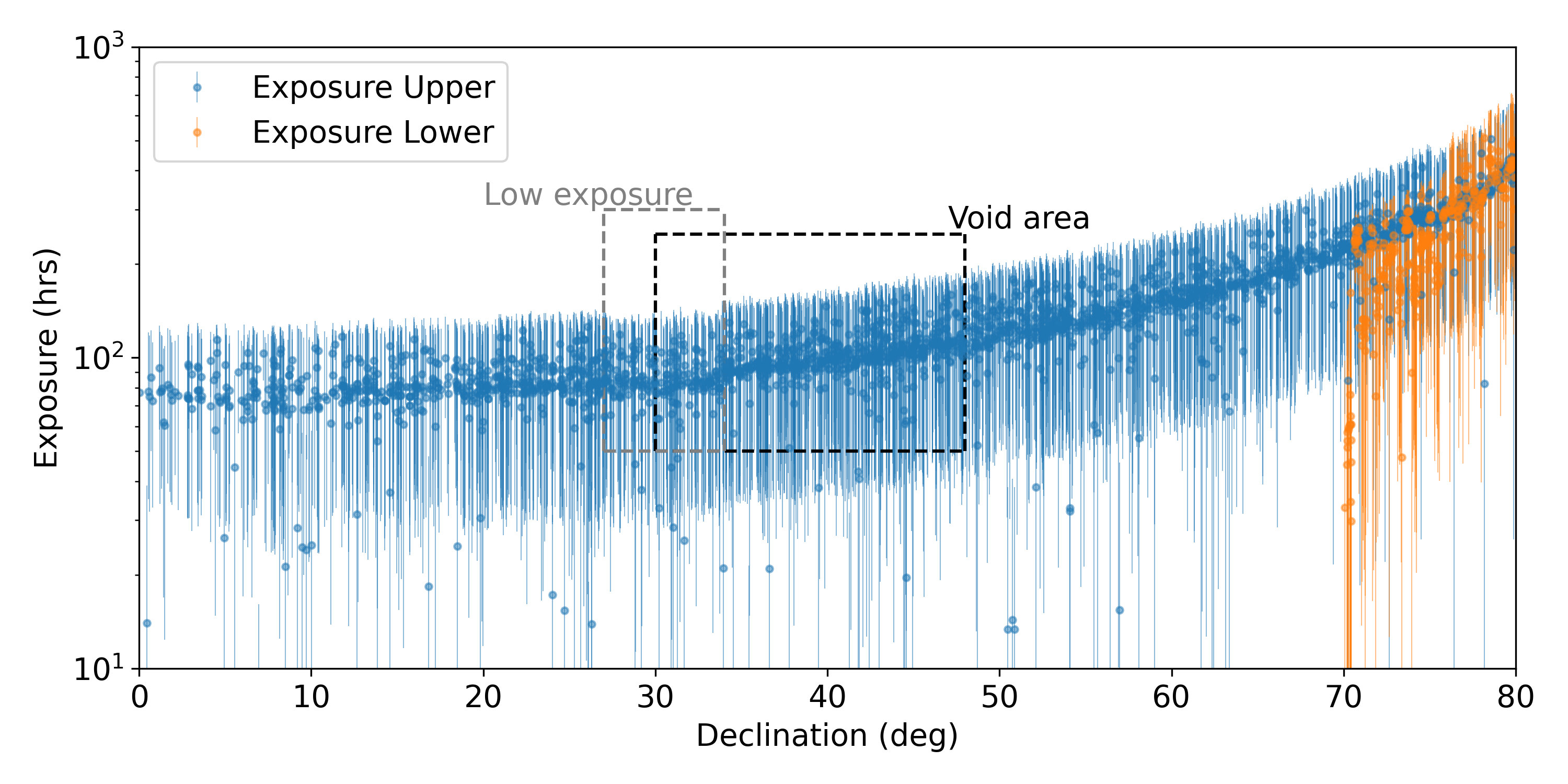}
    \caption{Exposure time (in hours) as a function of declination for both upper and lower transit beams. Low exposure region corresponding to a period of computational downtime is indicated by gray square and the detection gap area is highlighted by black square.}
    \label{fig:red-exp-dec}
\end{figure}

During part of the observing duration, there was reduced exposure between declinations $\sim27\degree$ and $\sim34\degree$ due to a period of computational downtime as shown in Figure \ref{fig:red-exp-dec}. A dip in exposure is present in the area enclosed by gray square. 

\bibliography{frbrefs, psrrefs, references}
\bibliographystyle{aasjournalv7}
\end{document}